\documentclass[12pt,english,aps,manuscript,showpacs]{revtex4-1}
\usepackage{graphicx}
\usepackage{dcolumn}
\usepackage{bm}
\usepackage{color}
\usepackage{hhline}
\usepackage{caption}
\usepackage{subcaption}

\begin{document}

\title{What is the orientation of the tip in a scanning tunneling microscope?}

\author{G\'abor M\'andi}
\address{
Department of Theoretical Physics, Budapest University of Technology and Economics,
Budafoki \'ut 8., H-1111 Budapest, Hungary
}
\author{Gilberto Teobaldi}
\address{
Stephenson Institute for Renewable Energy and Surface Science Research Centre, Department of Chemistry,
University of Liverpool, L69 3BX Liverpool, United Kingdom
}
\author{Kriszti\'an Palot\'as}
\email{palotas@phy.bme.hu}
\address{
Department of Theoretical Physics, Budapest University of Technology and Economics,
Budafoki \'ut 8., H-1111 Budapest, Hungary\\
}

\begin{abstract}

The atomic structure and electronic properties of the tip apex can strongly affect the contrast of
scanning tunneling microscopy (STM) images. This is a critical issue in STM imaging given the, to date unsolved, experimental
limitations in precise control of the tip apex atomic structure. Definition of statistically robust procedures to indirectly
obtain information on the tip apex structure is highly desirable as it would open up for more rigorous interpretation and
comparison of STM images from different experiments. To this end, here we introduce a statistical correlation analysis method to
obtain information on the local geometry and orientation of the tip used in STM experiments based on large scale
simulations. The key quantity is the relative brightness correlation of constant-current topographs between experimental and
simulated data. This correlation can be analyzed statistically for a large number of modeled tip orientations and
geometries. Assuming a stable tip during the STM scans and based on the correlation distribution, it is possible to determine
the tip orientations that are most likely present in an STM experiment, and exclude other orientations. This is
especially important for substrates such as highly oriented pyrolytic graphite (HOPG) since its STM contrast is strongly tip
dependent, which makes interpretation and comparison of STM images very challenging.
We illustrate the applicability of our method considering the HOPG surface in combination
with tungsten tip models of two different apex geometries and 18144 different orientations.
We calculate constant-current profiles along the $\langle 1\overline{1}00\rangle$ direction of the HOPG(0001)
surface in the $|V|\le 1$ V bias voltage range, and compare them with experimental data. We find that a blunt tip model provides
better correlation with the experiment for a wider range of tip orientations and bias voltages than a sharp tip model.
Such a combination of experiments and large scale simulations opens up the way for obtaining more detailed information on the
structure of the tip apex and more reliable interpretation of STM data in the view of local tip geometry effects.\\


\end{abstract}

\maketitle

\section{Introduction}

The interpretation of scanning tunneling microscopy (STM) images is not straightforward due to the effects of the local tip apex
geometry, termination and orientation. The reason is the convolution of sample and tip electronic states in a given
energy window defined by the bias voltage, and the fact that in STM experiments the detailed atomic geometry around the tip apex
is practically unknown and hardly controllable. On the other hand, it is clear that the electronic states and their dominating
orbital characters involved in the tunneling depend very much on the local atomic structure of the tip apex.

It has been a challenge to obtain information about the relevant properties of the STM tip apex for a long time.
Herz {\it et al.} performed reverse STM imaging experiments to study $p$, $d$, and $f$ orbital characters of the tip apex atom
above the Si(111)-($7\times 7$) surface \cite{herz03}. The combination of STM experiments and simulations on well characterized
surfaces to obtain information on the tip structure and termination was used, e.g., by Chaika {\it et al.}
\cite{chaika10,chaika13}. They considered the highly oriented pyrolytic graphite (HOPG) surface in the (0001) crystallographic
orientation in combination with W(001) tip models.
Rodary {\it et al.} studied Cr/W tip apex structures by high resolution transmission electron microscopy, and they pointed out
that the magnetization direction of monocrystalline nanotips cannot be controlled in spin-polarized STM \cite{rodary11}.
Recently, the effect of the tip orbitals on the STM imaging of supported molecular structures attracted considerable attention.
Gross {\it et al.} investigated pentacene and naphthalocyanine molecules on NaCl/Cu(111) surface by CO-functionalized tips,
and they explained the obtained STM contrast by tunneling through the $p$-states of the CO molecule \cite{gross11}.
Siegert {\it et al.} developed a reduced density matrix formalism in combination with Chen's derivative rule \cite{chen90} to
describe electron transport in STM junctions for molecular quantum dots, and studied the effect of selected tip orbital symmetries
on the STM images of the hydrogen phthalocyanine molecule on a thin insulating film \cite{siegert13}.
Lakin {\it et al.} proposed a method to deconvolute STM images and determine molecular orientations of both the sample and the
functionalized tip \cite{lakin13}.
In their work a C$_{60}$-Si(111)-($7\times 7$) surface and a C$_{60}$-functionalized tip were chosen.

Even in seemingly less complicated STM junctions, only a few theoretical works focused on the effect of the tip orientation on the
STM images. Hagelaar {\it et al.} demonstrated that a wide range of modeled tip terminations and orientations can reproduce the
experimental images for NO adsorbed on Rh(111) \cite{hagelaar08}. This work also showed that the modeling of realistic tip
structures, including nonsymmetric tips, is desirable for a good qualitative reproduction of experimental STM images. However,
it is quite unlikely that the relative orientation of the sample surface and the local tip apex geometry in STM experiments is of
high symmetry, which has been commonly assumed in the vast majority of STM simulations to date. M\'andi {\it et al.}
studied the effect of asymmetric relative tip-sample orientations on the STM contrast of the W(110) metal surface
\cite{mandi13tiprot} and of the HOPG(0001) surface \cite{mandi14rothopg} employing a three-dimensional Wentzel-Kramers-Brillouin
(3D-WKB) electron tunneling theory. It was found that the STM images can be substantially distorted due to tip geometry effects.
A physical explanation was provided based on the real-space shape of the electron orbitals entering the orbital-dependent
tunneling transmission formula in the 3D-WKB method \cite{mandi13tiprot}, see Eq.(\ref{Eq_Transmission}) in Appendix.
Motivated by the ideas of Hagelaar {\it et al.} and based on the methodology of M\'andi {\it et al.}, in the present work
a new concept of obtaining information about the local spatial orientation of the STM tip in real instruments is introduced.
The concept is substantiated by a combination of STM experiments and large scale simulations taking the HOPG(0001) surface.
Concomitantly, the qualitative visual analysis of STM images is advanced by quantifying their correspondence in terms of
relative brightness correlations.

The paper is organized as follows: The proposed correlation analysis method is introduced in section \ref{sec_method}, followed by
its application to the HOPG(0001) surface. We analyze and discuss our results in section \ref{sec_results}, and summarize our
findings in section \ref{sec_conclusions}.
The appendix reports a brief summary of the 3D-WKB tunneling theory with an arbitrary tip orientation.

\section{Method}
\label{sec_method}

To quantitatively compare the experimental (EXP) and simulated (SIM) constant-current topographs, the definition of the
relative brightness of a given two-dimensional (2D) contour $C$ at bias voltage $V_k$ is needed \cite{mandi14rothopg,teobaldi12}:
\begin{equation}
\label{Eq_brightness}
B_{C}(\mathbf{x},V_k)=\frac{z_{C}(\mathbf{x},V_k)-z_{C}(\mathbf{x}_{\mathrm{min}},V_k)}{z_{C}(\mathbf{x}_{\mathrm{max}},V_k)-z_{C}(\mathbf{x}_{\mathrm{min}},V_k)},
\end{equation}
where $z_{C}(\mathbf{x},V_k)$ is the apparent height of the constant-current contour $C$ above the surface lateral
$\mathbf{x}=x_{ij}$ position at bias voltage $V_k$ obtained by $C\in\{\mathrm{EXP,SIM}\}$. $z_{C}(\mathbf{x}_{\mathrm{min}},V_k)$
and $z_{C}(\mathbf{x}_{\mathrm{max}},V_k)$ respectively have the smallest and largest apparent heights in the 2D scan area, thus
due to the definition, $B_{C}(\mathbf{x}_{\mathrm{min}},V_k)=0$ and $B_{C}(\mathbf{x}_{\mathrm{max}},V_k)=1$. Assuming that all
$B_{C}(x_{ij},V_k)$ contours consist of $N_x\times N_y$ points ($i=1,...,N_x$, $j=1,...,N_y$), the mean value of the
relative brightness in a given bias voltage range of $N_V$ bias values ($k=1,...,N_V$) can be calculated as
\begin{equation}
\label{Eq_meanbrightness}
\overline{B}_{C}=\frac{1}{N_x N_y N_V}\sum_{k=1}^{N_V}\sum_{i=1}^{N_x}\sum_{j=1}^{N_y}B_{C}(x_{ij},V_k).
\end{equation}
Using the same resolution of the scanning area in the experiment and in the simulations resulting in relative brightness contours
of $N_x\times N_y$ lateral points in both cases, it is possible to quantitatively compare the $B_{\mathrm{EXP}}$ and
$B_{\mathrm{SIM}}$ contours in the corresponding bias voltage range of $N_V$ bias values by calculating their
correlation coefficient as
\begin{eqnarray}
r&=&\left\{\sum_{k=1}^{N_V}\sum_{i=1}^{N_x}\sum_{j=1}^{N_y}[B_{\mathrm{EXP}}(x_{ij},V_k)-\overline{B}_{\mathrm{EXP}}][B_{\mathrm{SIM}}(x_{ij},V_k)-\overline{B}_{\mathrm{SIM}}]\right\}\nonumber\\
&\times&\left\{\sum_{k=1}^{N_V}\sum_{i=1}^{N_x}\sum_{j=1}^{N_y}[B_{\mathrm{EXP}}(x_{ij},V_k)-\overline{B}_{\mathrm{EXP}}]^2\right\}^{-1/2}\nonumber\\
&\times&\left\{\sum_{k=1}^{N_V}\sum_{i=1}^{N_x}\sum_{j=1}^{N_y}[B_{\mathrm{SIM}}(x_{ij},V_k)-\overline{B}_{\mathrm{SIM}}]^2\right\}^{-1/2}.
\label{Eq_correlation}
\end{eqnarray}
The (Pearson product-moment) correlation coefficient $r$ measures the degree of linear relationship between the
$B_{\mathrm{EXP}}(x_{ij},V_k)$ and $B_{\mathrm{SIM}}(x_{ij},V_k)$ datasets. Due to the definition, the values of $r$ are bounded
to the range of [-1, +1]. $r=+1$ corresponds to a perfect positive linear relationship that is desirable when comparing relative
brightness contours between experiment and simulations. Obtaining $r=+1$ would mean that the simulation reproduces the
experimental data perfectly. $r=-1$ means a perfect negative linear relationship, e.g., this would be the result of calculating
the correlation coefficient of exactly oppositely corrugated contours.
$r=0$ means that there is no linear relationship between the contours.

Another statistical measure for the difference between experimental and simulated contours is the mean squared error,\\
MSE=$\frac{1}{N_x N_y N_V}\sum_{k=1}^{N_V}\sum_{i=1}^{N_x}\sum_{j=1}^{N_y}[B_{\mathrm{EXP}}(x_{ij},V_k)-B_{\mathrm{SIM}}(x_{ij},V_k)]^2$.\\
A perfect agreement of contours is obtained at MSE=0, and it is desired that MSE is minimal comparing experimental and
simulated contours for obtaining the best agreement. For selected contours and bias voltages we found good correspondence between
minimal MSE and maximal correlation. However, MSE is not bounded from above, and this makes the analysis of MSE distribution and
the interpretation of maximal MSE difficult. Therefore, we excluded using this measure in our statistical analysis.

The calculation of the correlation coefficient in Eq.(\ref{Eq_correlation}) was presented in the more general case of taking
2D relative brightness contours. However, the same method can be specifically applied to one-dimensional (1D)
relative brightness profiles by setting $N_y=1$. This approach will be used in the paper for the $\langle 1\bar{1}00\rangle$
direction of the HOPG(0001) surface since experimental data \cite{teobaldi12} is available for such a case.
To calculate the relative brightness correlations between the experiment and simulations, profiles shifted to start with
their minimum value, $B_{C}(x_{i=1,j=1},V_k)=0$ are taken. A detailed discussion justifying this was given in section 3.2.\ of
Ref.\ \cite{mandi14rothopg}.

Since in the simulations the tip material (TIPMAT), atomic arrangement/geometry (TIPGEO), and orientation described by the
Euler angles $(\theta_0,\phi_0,\psi_0)$
can be chosen in practically infinite ways, the corresponding relative brightness profiles are dependent on these parameters:\\
$B_{SIM}(\mathbf{x},V_k)=B_{SIM}(\mathbf{x},V_k,\mathrm{TIPMAT_{TIPGEO}},\theta_0,\phi_0,\psi_0)$, and similarly,
the correlation coefficient is $r=r(\mathrm{TIPMAT_{TIPGEO}},\theta_0,\phi_0,\psi_0)$.
In the present work, we consider TIPMAT=W (tungsten) and TIPGEO $\in\{\mathrm{blunt,sharp}\}$ tip models.
The $\mathrm{W_{blunt}}$ tip is represented by an adatom adsorbed on the hollow site of the W(110) surface and
the $\mathrm{W_{sharp}}$ tip is modeled as a pyramid of three-atoms height on the W(110) surface.
More details on the used tip geometries can be found in Ref.\ \cite{teobaldi12}.
These tip models are expected to bracket the range of possible tip sharpnesses in experiments as extremely blunt tips with flat
surfaces would provide no contrast-resolution at all and sharp pyramids would likely be very unstable during prolonged tip scans.
Moreover, carbon-contaminated tips with a C atom at the apex can be excluded due to a dramatic decrease of the tunneling current
\cite{teobaldi12}.

In our simulations the three-dimensional Wentzel-Kramers-Brillouin (3D-WKB) electron tunneling theory with arbitrary tip
orientations is employed, see Appendix, which is implemented in the 3D-WKB-STM code
\cite{mandi13tiprot,palotas12orb,palotas13fop}. Recently, the 3D-WKB method was successfully applied in a number of theoretical
\cite{palotas11sts,palotas11stm,palotas12sts,palotas13contrast,mandi14fe} and combined experimental-theoretical investigations
\cite{nita14prb}.

\begin{figure*}
\includegraphics[width=0.67\textwidth,angle=0]{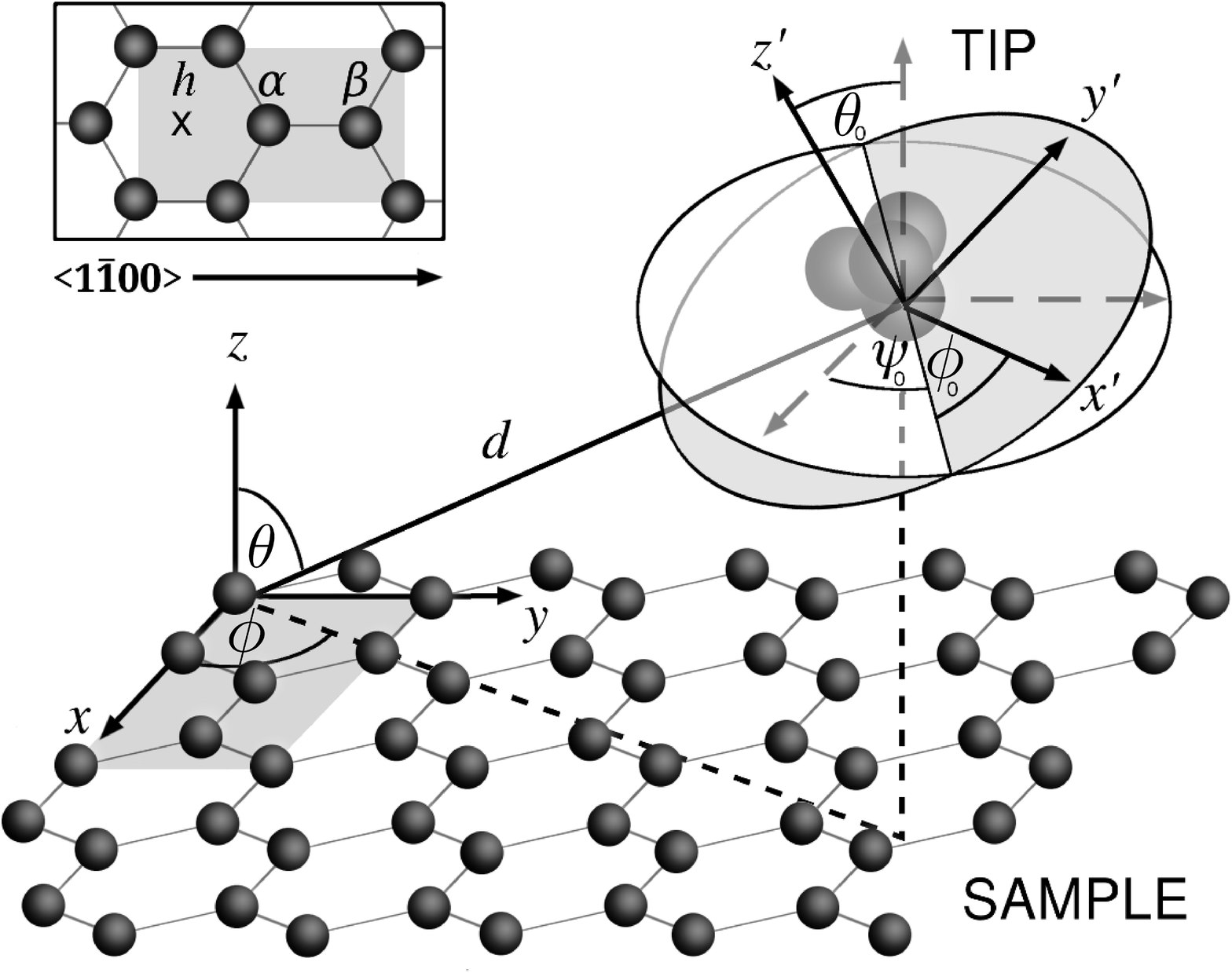}
\caption{\label{Fig1} Schematic view of the STM tip above the HOPG surface. The rotation of the local coordinate
system of the tip with respect to that of the sample surface is described by the Euler angles $(\theta_0,\phi_0,\psi_0)$.
Inset shows the positions of the characteristic $h$, $\alpha$ and $\beta$ sites of the HOPG(0001) surface
along the $\langle 1\bar{1}00\rangle$ direction.
}
\end{figure*}

Constant-current brightness profiles are calculated along the $\langle 1\bar{1}00\rangle$ direction
($x$-axis of Fig.\ \ref{Fig1}) containing the three characteristic positions of the HOPG(0001) surface: hollow ($h$),
$\alpha$-carbon and $\beta$-carbon, see inset of Fig.\ \ref{Fig1}. The experimental averaged brightness data with
$N_y=1$ and $N_x=46$ points are taken from Fig.\ 4 of Ref.\ \cite{teobaldi12} in the interval of [-1 V, 1 V]
with 0.1 V steps. In the simulations the current values are chosen for each corresponding bias voltage in such a way that the
lowest apparent height of each constant-current contour is $z_{\mathrm{SIM}}(\mathbf{x}_{\mathrm{min}},V_k)=5.5$ \AA$\;$
(pure tunneling regime). The relative brightness profiles are calculated by using the introduced
$\mathrm{W_{blunt}}$ and $\mathrm{W_{sharp}}$ tip models for a set of tip orientations described by the Euler angles:
$\theta_0\in [0^{\circ},30^{\circ}]$, $\phi_0\in [0^{\circ},175^{\circ}]$, $\psi_0\in [0^{\circ},355^{\circ}]$
with $5^{\circ}$ steps. The Euler angles are visualized in Fig.\ \ref{Fig1}.
$\theta_0$ angle describes the rotation with respect to the $x$ axis, transforming the $z$ axis to $z'$.
Additionally, $\phi_0$ and $\psi_0$ are rotation angles around the $z'$ and $z$ axes, respectively,
as Fig.\ \ref{Fig1} shows. The exact meaning of the Euler angles is mathematically formulated in the rotation matrix in
Eq.(\ref{Eq_mrot}) in Appendix and explained in Refs.\ \cite{mandi13tiprot,mandi14rothopg}.
Altogether $7\times 36\times 72=18144$ tip orientations are considered.
For this selection we used the general symmetry property of the rotation matrix in Eq.(\ref{Eq_mrot}):
$(\theta_0,\phi_0,\psi_0)=(-\theta_0,\phi_0+\pi,\psi_0+\pi)$ and the mirror symmetry of the HOPG surface
above the $h-\alpha-\beta$ line: $(\theta_0,\phi_0,\psi_0)=(-\theta_0,-\phi_0,-\psi_0)$.
Correlation coefficients in Eq.(\ref{Eq_correlation}) are calculated between the experimental
and a large number of simulated relative brightness profiles in the negative (-1 V $\le V<0$ V, $N_V=10$),
positive (0 V $<V\le$ 1 V, $N_V=10$) and full (-1 V $\le V\le$ 1 V, $N_V=20$) bias voltage ranges.

\section{Results and discussion}
\label{sec_results}

\begin{figure*}
\begin{subfigure}[h]{0.32\textwidth}
\includegraphics[width=1.0\textwidth,angle=0]{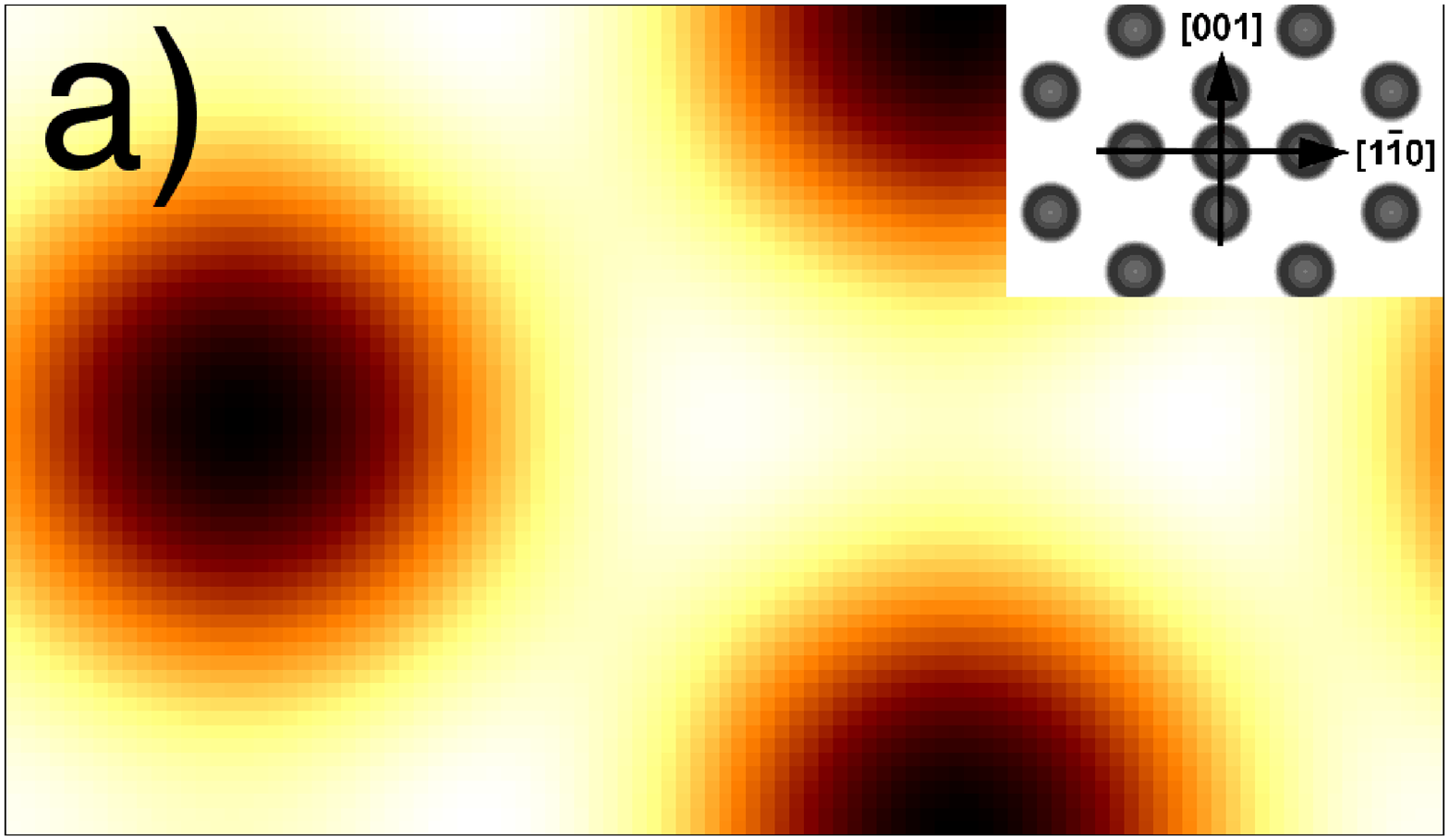}
\end{subfigure}
\begin{subfigure}[h]{0.32\textwidth}
\includegraphics[width=1.0\textwidth,angle=0]{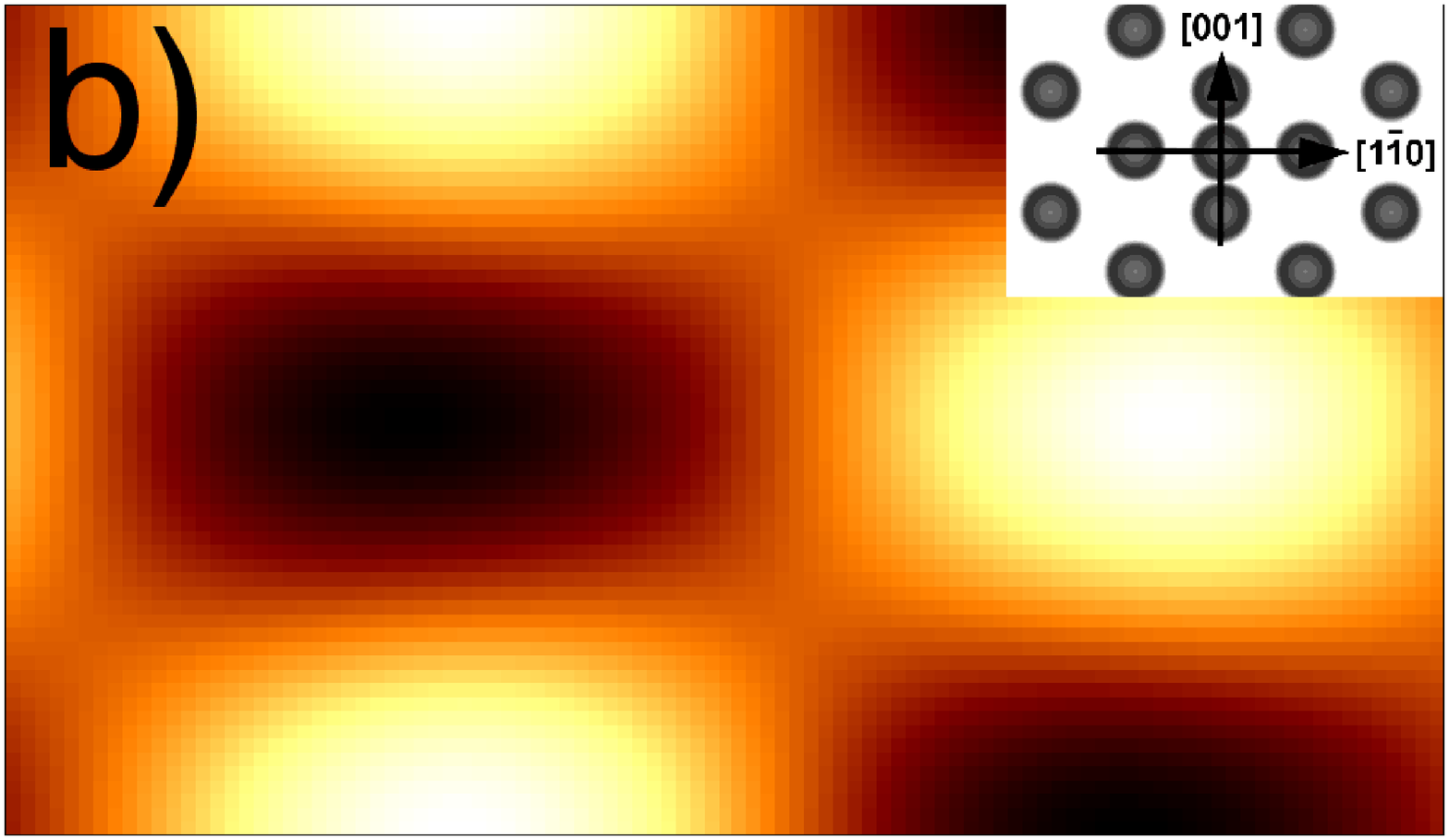}
\end{subfigure}
\begin{subfigure}[h]{0.32\textwidth}
\includegraphics[width=1.0\textwidth,angle=0]{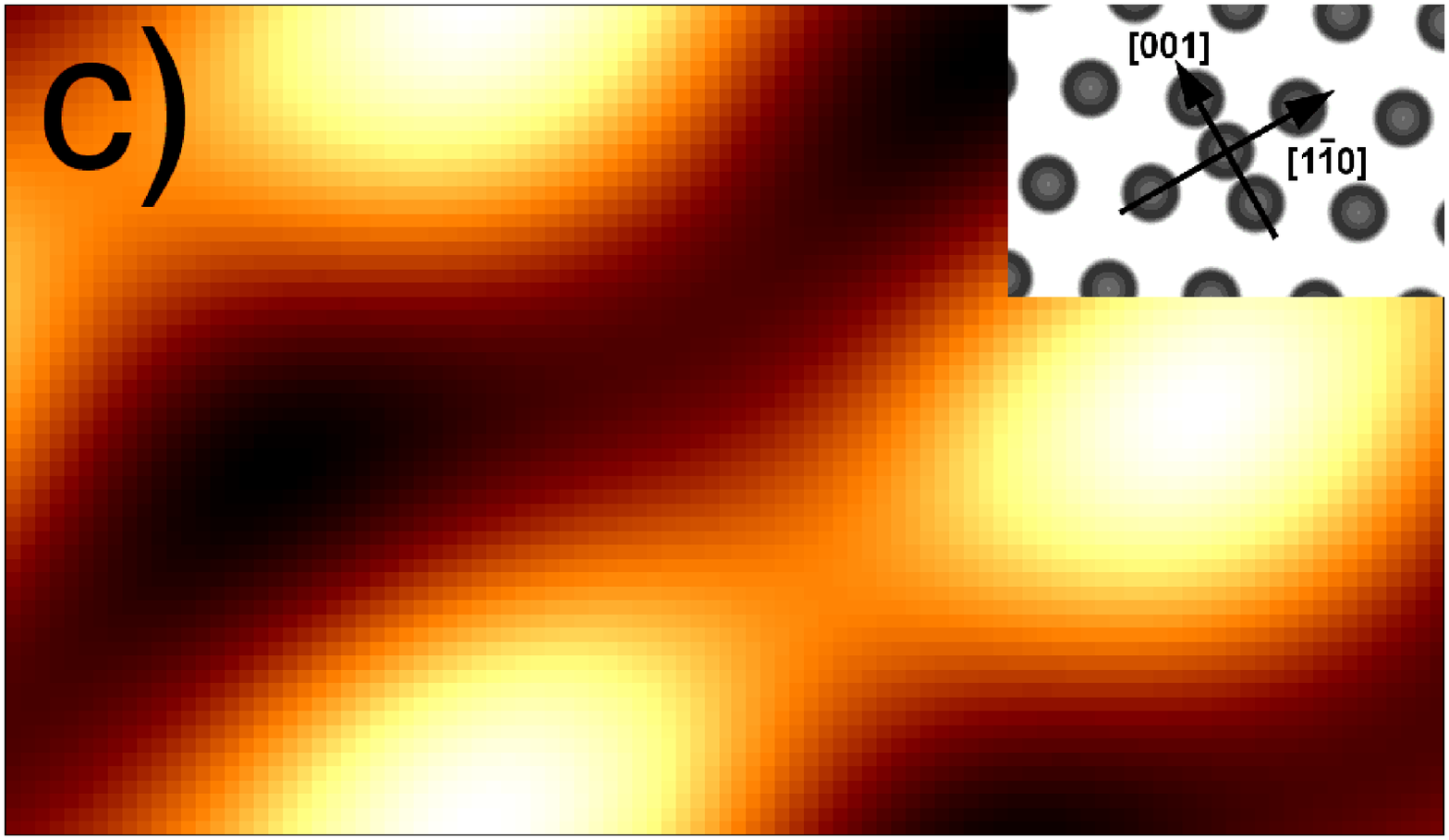}
\end{subfigure}
\caption{\label{Fig2} Constant-current STM images illustrating the variety of observed STM contrasts above the HOPG(0001) surface
in the tunneling regime for $\theta_0=\phi_0=0^{\circ}$:
a) hexagonal contrast (both $\alpha$- and $\beta$-carbons are bright; $V=1$ V, $\psi_0=90^{\circ}$),
b) triangular contrast (only $\beta$-carbons are bright; $V=0.1$ V, $\psi_0=90^{\circ}$),
c) triangular contrast with striped feature ($V=0.1$ V, $\psi_0=120^{\circ}$).
The STM images are calculated above the shaded rectangular area shown in the inset of Fig.\ \ref{Fig1} using the
$\mathrm{W_{blunt}}$ tip model. Inset shows the relative orientation of the $\mathrm{W_{blunt}}$ tip
with respect to the HOPG(0001) surface in each subfigure.
}
\end{figure*}

We recall that the STM contrast of the HOPG(0001) surface can change substantially depending on the tunneling and
tip parameters \cite{chaika10,chaika13,mandi14rothopg,teobaldi12,ondracek11}. A selection of the possible STM contrasts in
the tunneling regime is shown in Fig.\ \ref{Fig2}. Here, the two nonequivalent carbon atoms of HOPG ($\alpha$ and $\beta$)
are primarily responsible for the different STM contrasts [hexagonal contrast in Fig.\ \ref{Fig2}a) and triangular contrast in
Fig.\ \ref{Fig2}b)]. Particular rotations of the STM tip were shown to result in striped STM images \cite{mandi14rothopg},
affecting the secondary contrast features [Fig.\ \ref{Fig2}c)]. In the near contact regime multiple scattering effects and
tip-sample forces also play an important role in the STM contrast appearance \cite{blanco04},
e.g., a shift of the maximum brightness from
the $\beta$-carbon to the hollow ($h$) position of HOPG was demonstrated by Ondr\'a\v{c}ek {\it {et al.}}
\cite{ondracek11}. Note that we restrict our study to the pure tunneling regime corresponding to the
used experimental data \cite{teobaldi12} and to the validity of the 3D-WKB method \cite{mandi14rothopg}.
The diversity of the observed STM contrasts above the HOPG(0001) surface surely contains information about the local geometry of
the tip apex in STM measurements, therefore HOPG(0001) is an ideal candidate to illustrate the applicability of our statistical
correlation analysis method combining large scale STM simulations with experiments.

\begin{figure*}
\begin{subfigure}[h]{0.32\textwidth}
\includegraphics[width=1.0\textwidth,angle=0]{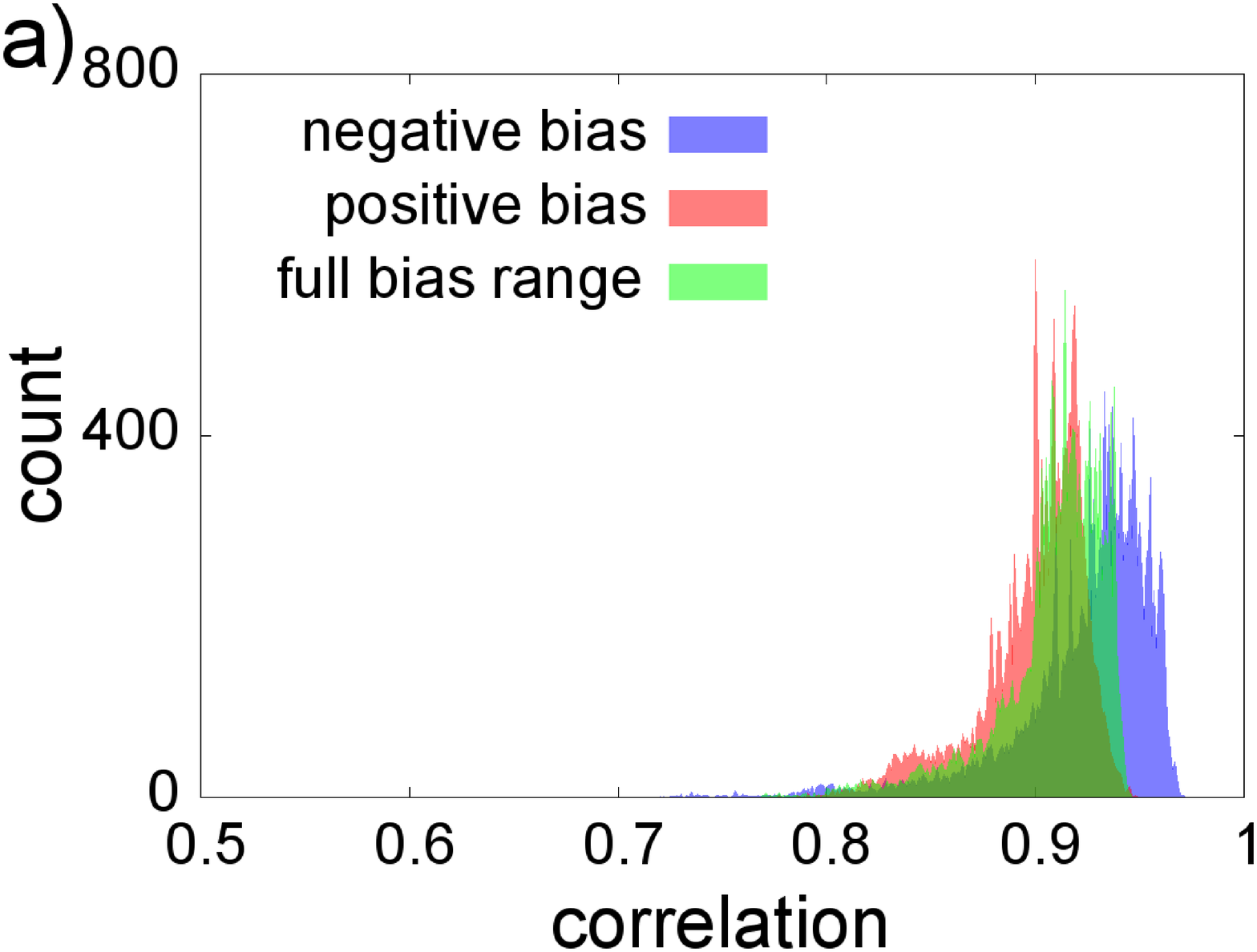}
\end{subfigure}
\begin{subfigure}[h]{0.32\textwidth}
\includegraphics[width=1.0\textwidth,angle=0]{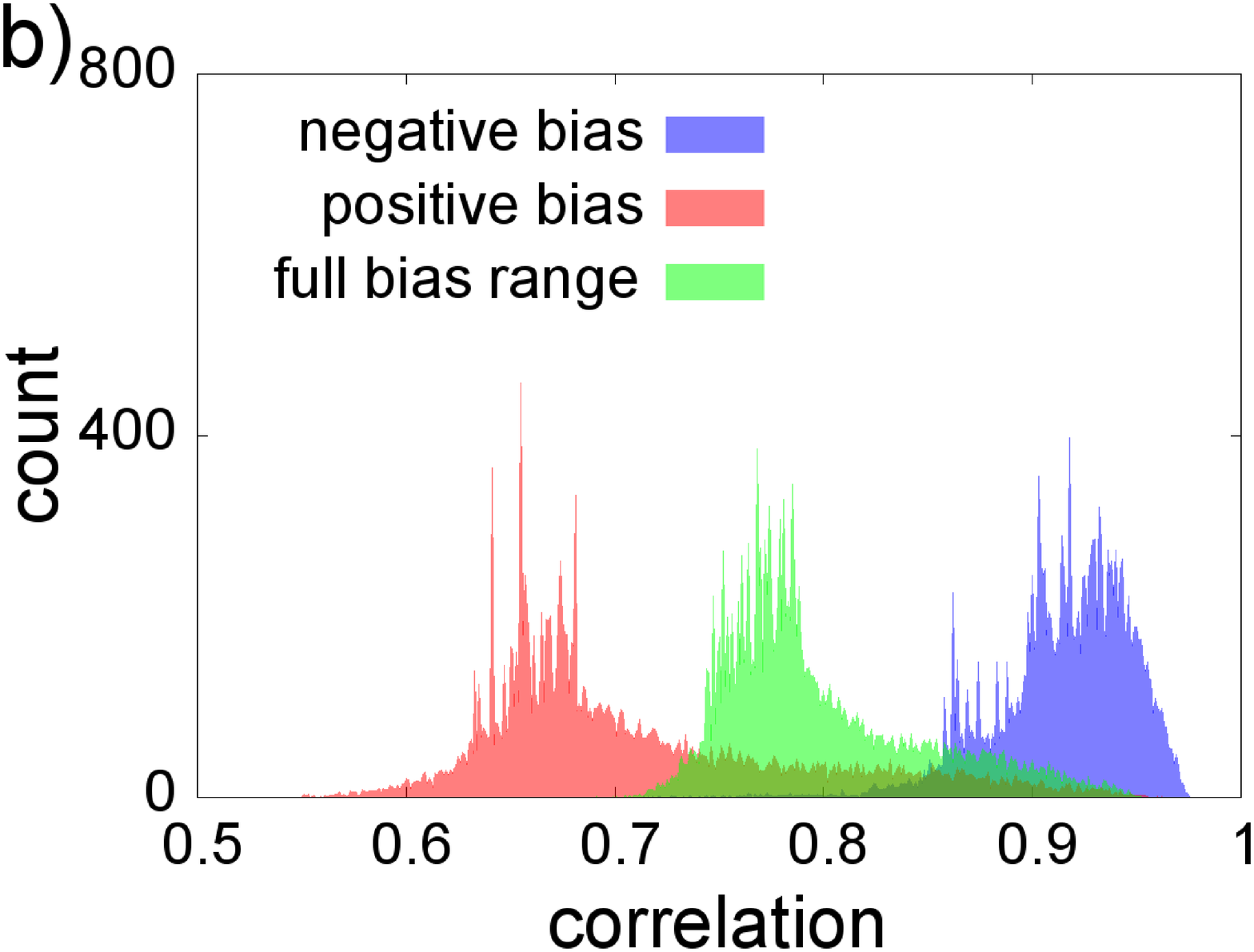}
\end{subfigure}
\begin{subfigure}[h]{0.32\textwidth}
\includegraphics[width=1.0\textwidth,angle=0]{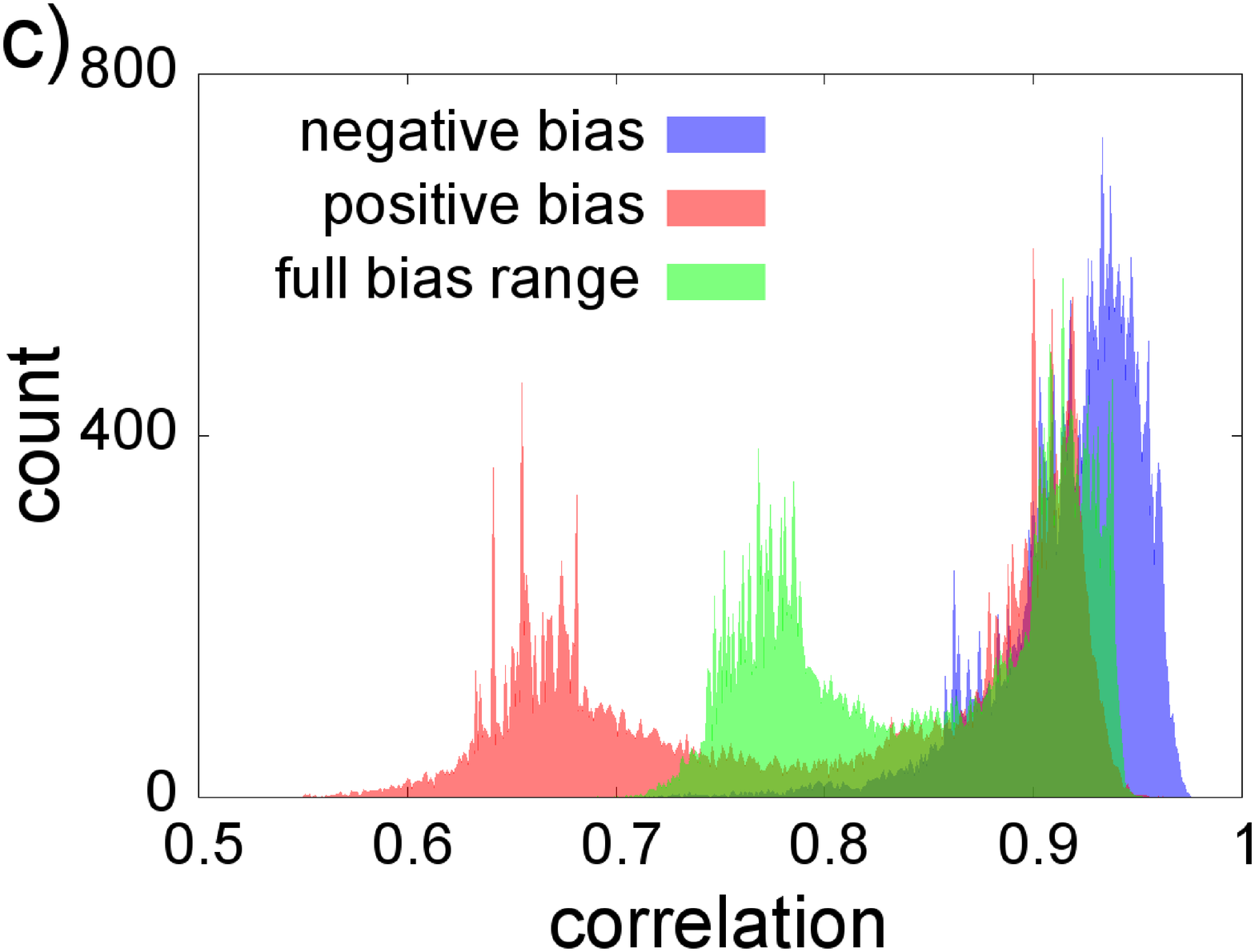}
\end{subfigure}
\caption{\label{Fig3} $|V|\le 1$ V relative brightness correlation histograms calculated by using 18144 tip orientations for:
a) $\mathrm{W_{blunt}}$ tip, b) $\mathrm{W_{sharp}}$ tip. Part c) reports the sum of the histograms in a) and b).
The correlation histograms for the negative, positive and full bias ranges are shown using Eq.(\ref{Eq_correlation}) in the
[0.5, 1] range with 0.001 resolution.
}
\end{figure*}

Fig.\ \ref{Fig3} shows the calculated relative brightness correlation histograms for the two considered tungsten tip models in
18144 tip orientations and the sum of the two histograms in Fig.\ \ref{Fig3}c). The maximal correlation between the experiment
and simulations is found at approximately 0.97 in the negative and at approx.\ 0.95 in the positive bias range for both tips.
However, we cannot conclude that the tip orientations belonging to the maximal correlation are the best since there is a large
number of other orientations within a few percent from the maximum correlation well above 0.9. Analyzing the correlation
distribution, it is clearly seen that much more tip orientations provide better correlation values in the negative compared to
the positive bias range for both tip models. This effect is even more evident for the $\mathrm{W_{sharp}}$ tip, where the
correlation distributions have two distinct peaks for the negative and positive bias at around 0.93 and 0.66, respectively.
The presented statistics for the
relative brightness correlation taking a large number of tip orientations confirm the significance of the findings of
Ref.\ \cite{mandi14rothopg}, where the simulated brightness profiles obtained at positive bias for the $\mathrm{W_{sharp}}$ tip
model in high symmetry orientations resulted in much lower correlation with the experiment than in the negative bias voltage
range. No such large differences were found for the $\mathrm{W_{blunt}}$ tip at either bias polarities. This suggests that the
$\mathrm{W_{blunt}}$ tip is more likely to be present in a wide range of bias voltages in the experiment than the
$\mathrm{W_{sharp}}$ tip.

The minimal correlation between the experimental and simulated brightness profiles is found at 0.55
for the $\mathrm{W_{sharp}}$ tip at positive bias voltages, whereas for the $\mathrm{W_{sharp}}$ tip at negative bias voltages
and for the $\mathrm{W_{blunt}}$ tip at all considered bias voltage ranges the minimal correlation is above 0.7. Once more, this
suggests a more likely $\mathrm{W_{blunt}}$ than $\mathrm{W_{sharp}}$ tip in the experiment since various local rotations of
the $\mathrm{W_{blunt}}$ tip do not give worse correlations with the experiment than 0.7, whereas
there are particular local rotations of the $\mathrm{W_{sharp}}$ tip at positive bias voltages with much worse correlations.

The presented relative brightness correlation histograms provide information about the distribution of the correlation values
in terms of the number of simulated tip orientations within a particular correlation range with the experimental brightness data.
This presentation of the correlation statistics, however, cannot tell which specific tip orientations give the best or worst
correlations with the experiment. To assign the most or least likely orientations of the STM tip in the experiment for the given
tip model, we need another representation of the correlation data. Therefore, we complement our analysis by calculating
correlation maps: $r(\mathrm{W_{blunt}},\theta_0,\phi_0,\psi_0)$ and $r(\mathrm{W_{sharp}},\theta_0,\phi_0,\psi_0)$.

\begin{figure*}
\begin{subfigure}[h]{0.42\textwidth}
\includegraphics[width=1.0\textwidth,angle=0]{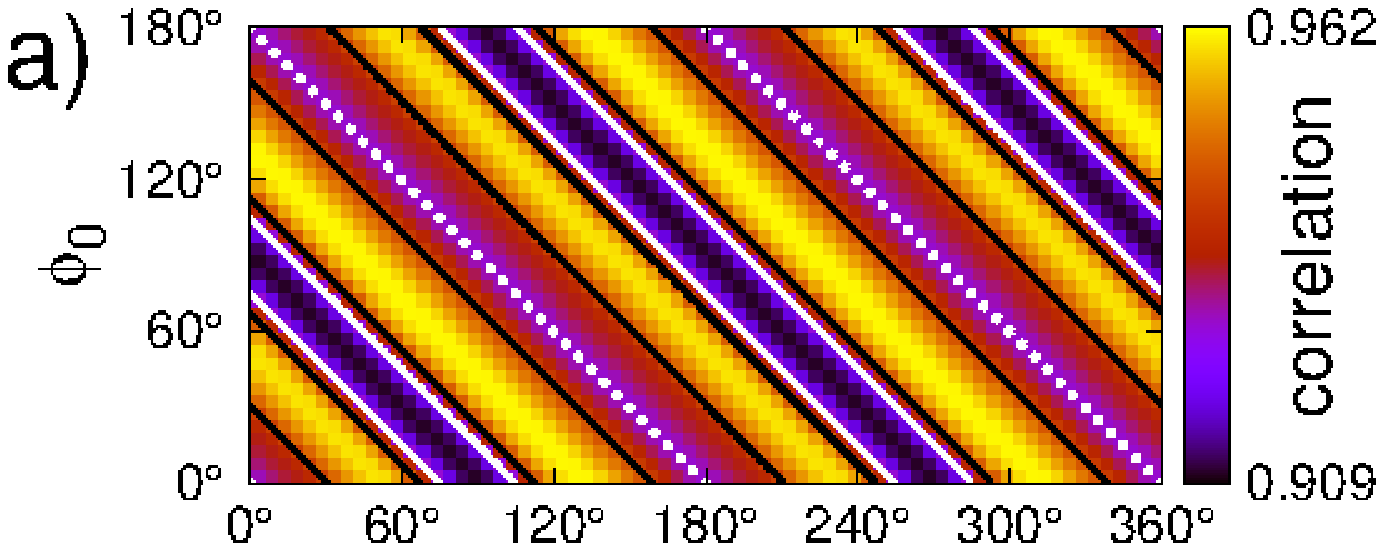}
\end{subfigure}
\begin{subfigure}[h]{0.42\textwidth}
\includegraphics[width=1.0\textwidth,angle=0]{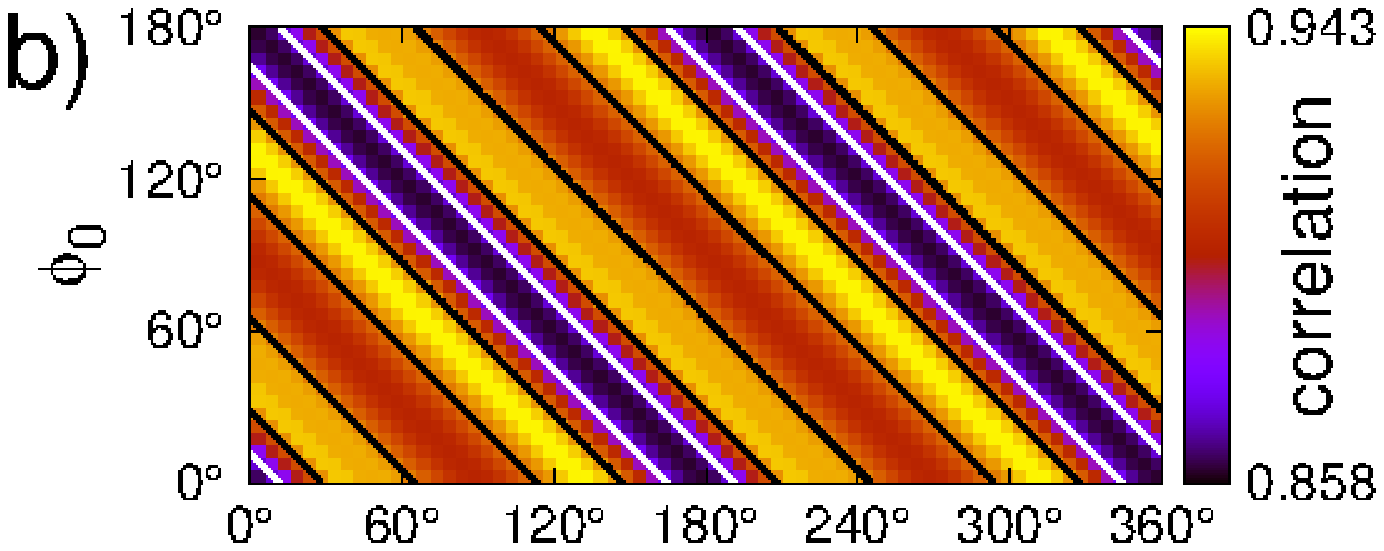}
\end{subfigure}

\begin{subfigure}[h]{0.42\textwidth}
\includegraphics[width=1.0\textwidth,angle=0]{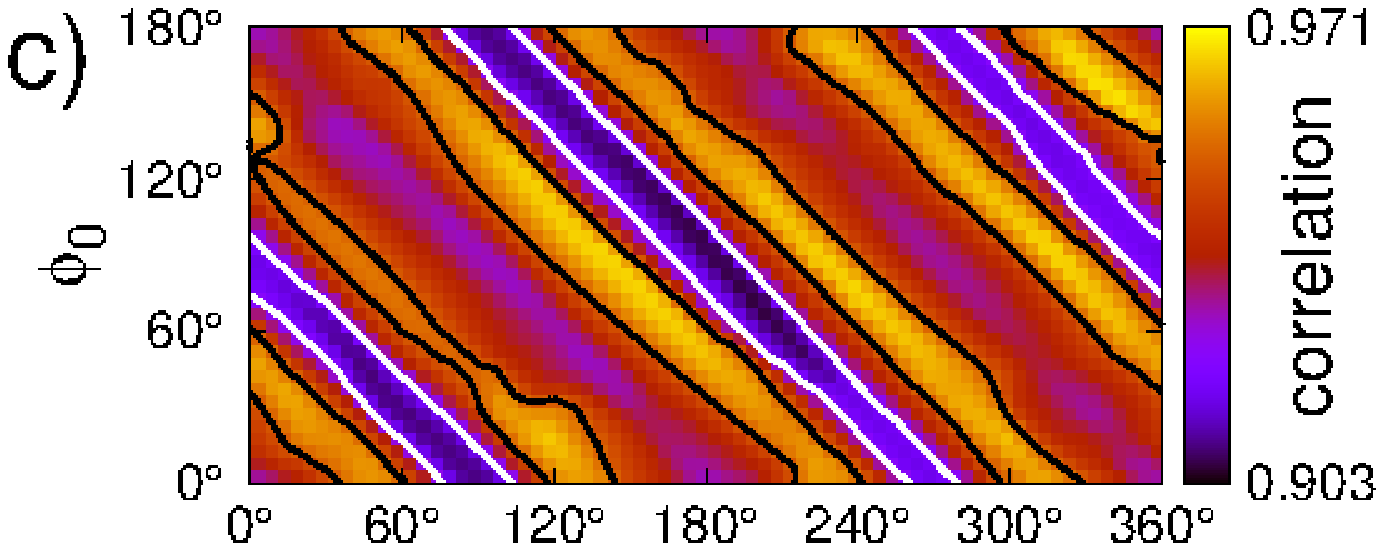}
\end{subfigure}
\begin{subfigure}[h]{0.42\textwidth}
\includegraphics[width=1.0\textwidth,angle=0]{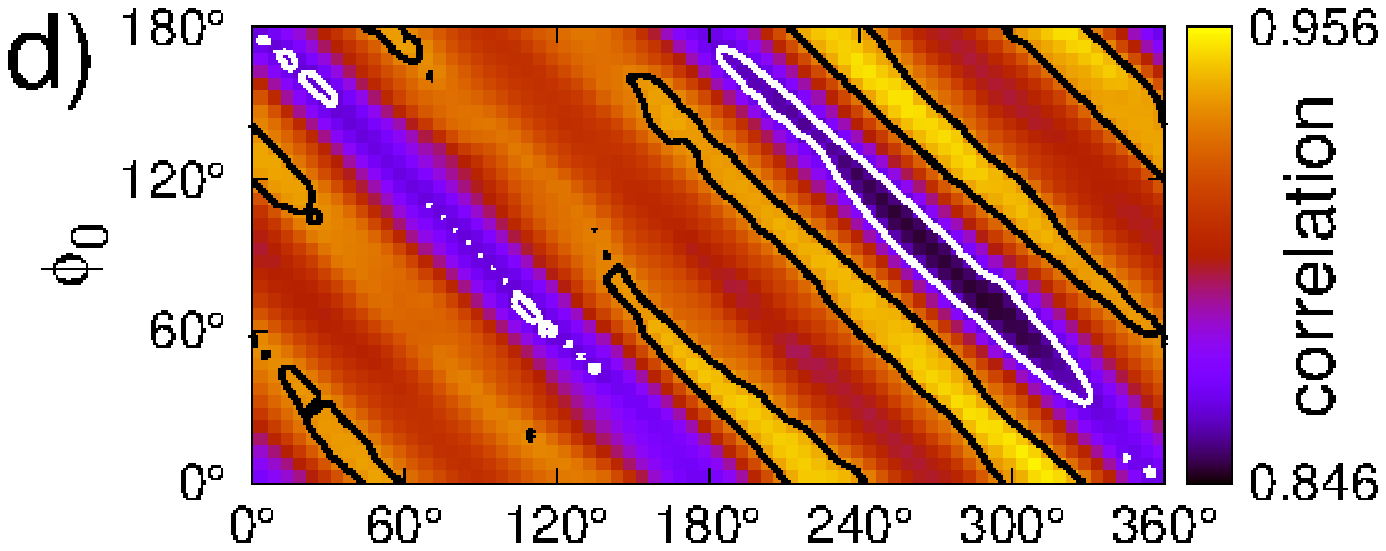}
\end{subfigure}

\begin{subfigure}[h]{0.42\textwidth}
\includegraphics[width=1.0\textwidth,angle=0]{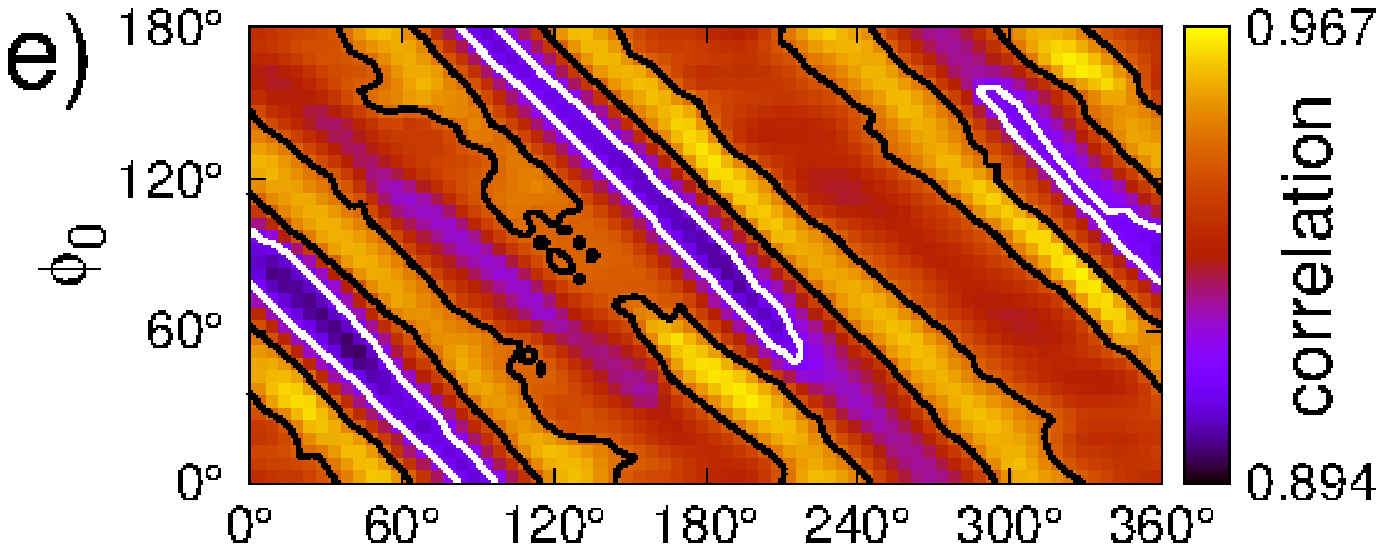}
\end{subfigure}
\begin{subfigure}[h]{0.42\textwidth}
\includegraphics[width=1.0\textwidth,angle=0]{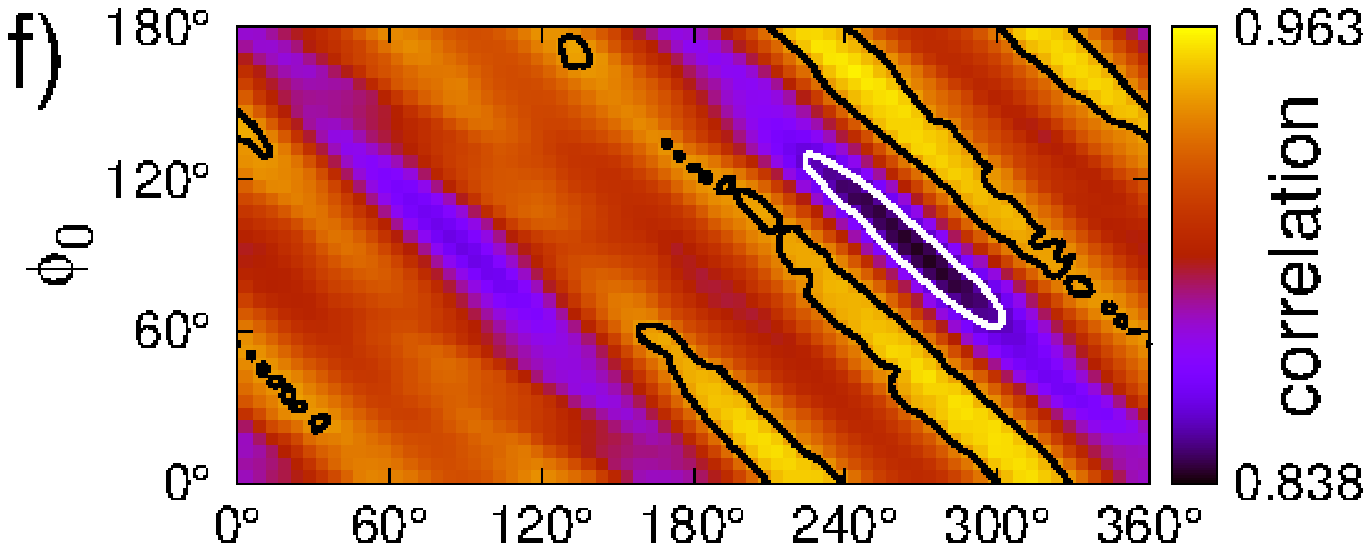}
\end{subfigure}

\begin{subfigure}[h]{0.42\textwidth}
\includegraphics[width=1.0\textwidth,angle=0]{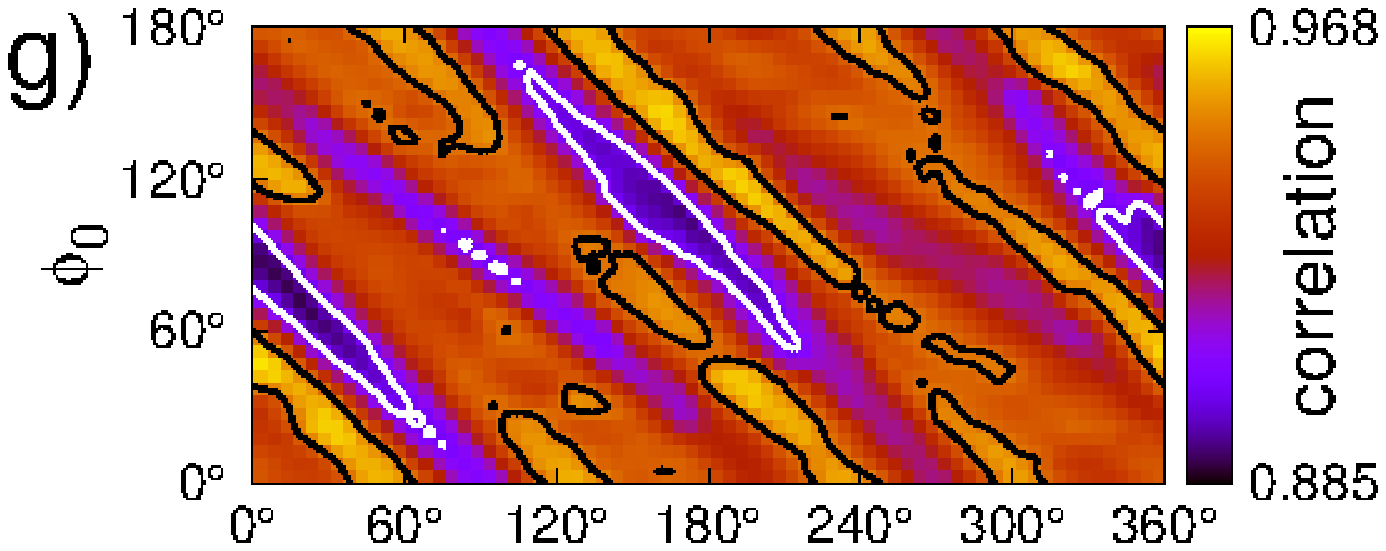} 
\end{subfigure}
\begin{subfigure}[h]{0.42\textwidth}
\includegraphics[width=1.0\textwidth,angle=0]{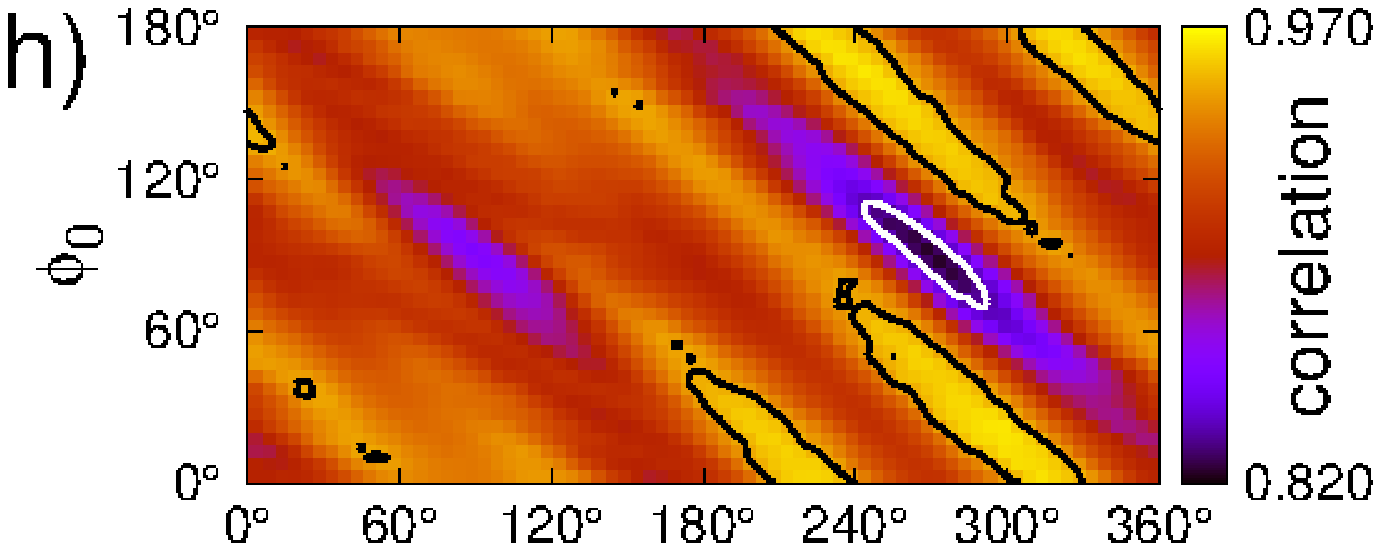}
\end{subfigure}

\begin{subfigure}[h]{0.42\textwidth}
\includegraphics[width=1.0\textwidth,angle=0]{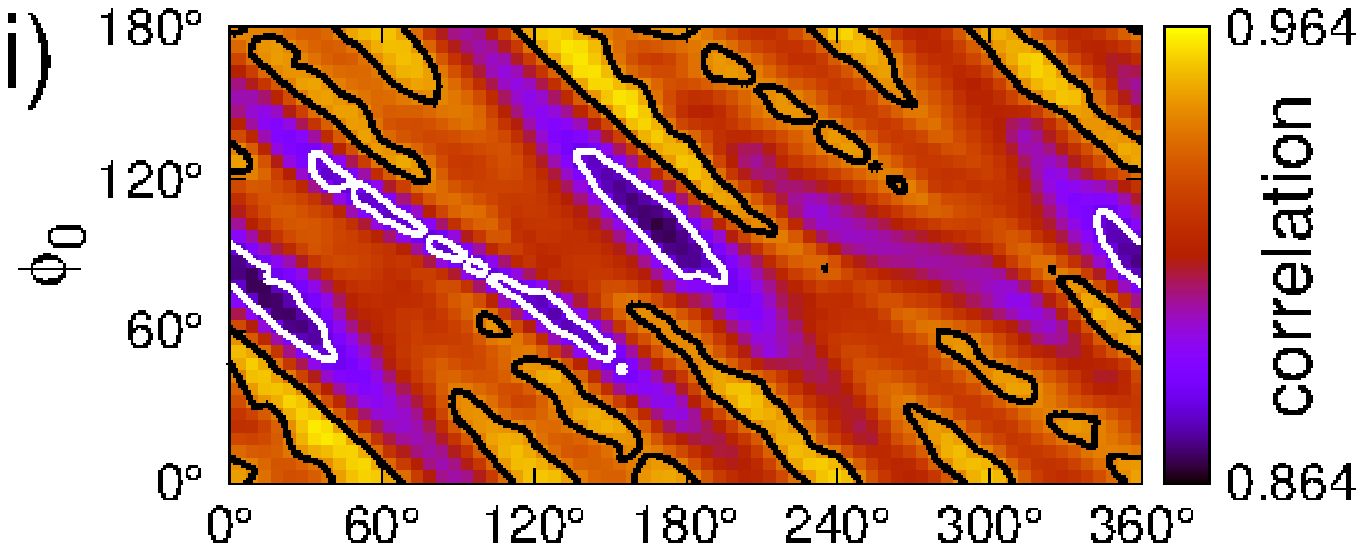}
\end{subfigure}
\begin{subfigure}[h]{0.42\textwidth}
\includegraphics[width=1.0\textwidth,angle=0]{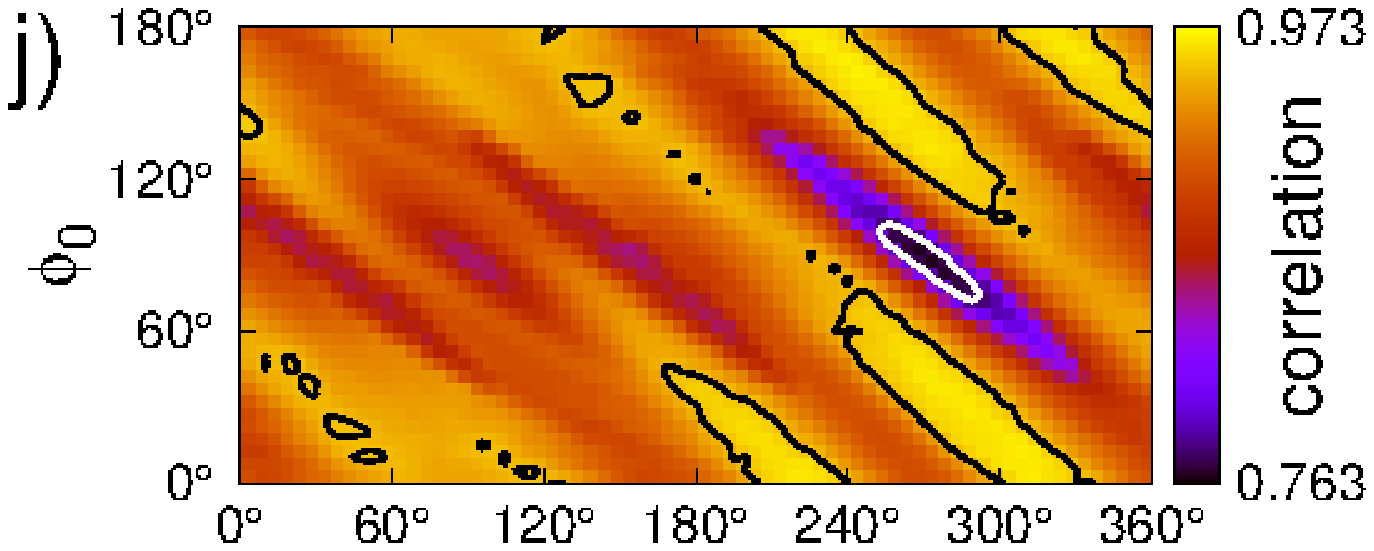}
\end{subfigure}

\begin{subfigure}[h]{0.42\textwidth} 
\includegraphics[width=1.0\textwidth,angle=0]{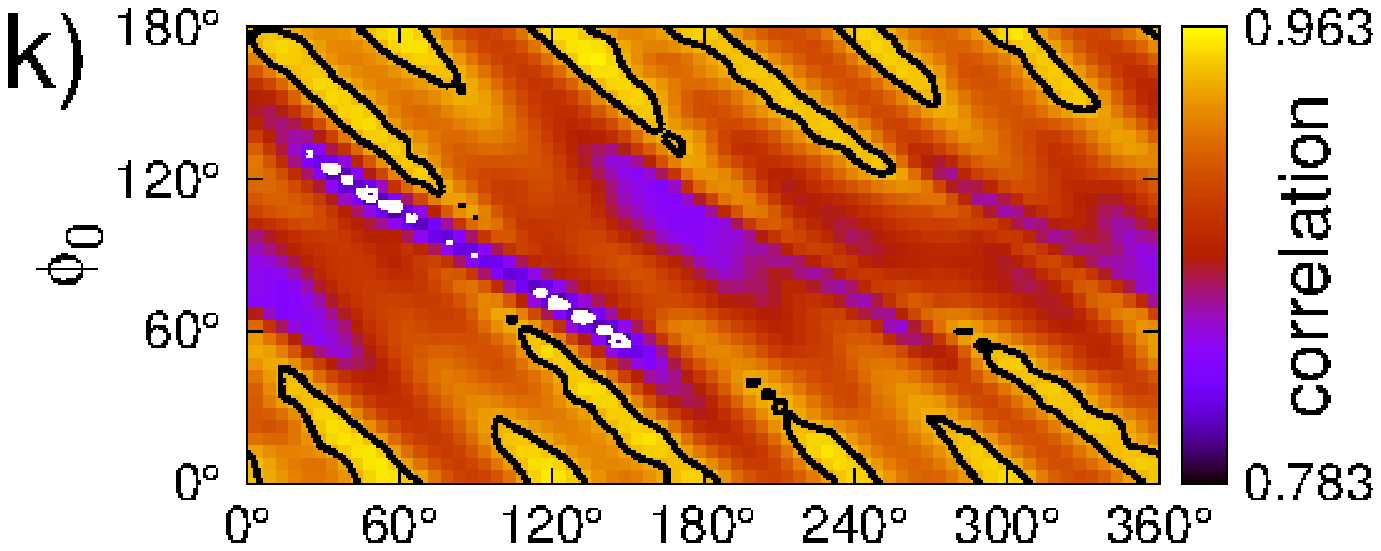}
\end{subfigure}
\begin{subfigure}[h]{0.42\textwidth}
\includegraphics[width=1.0\textwidth,angle=0]{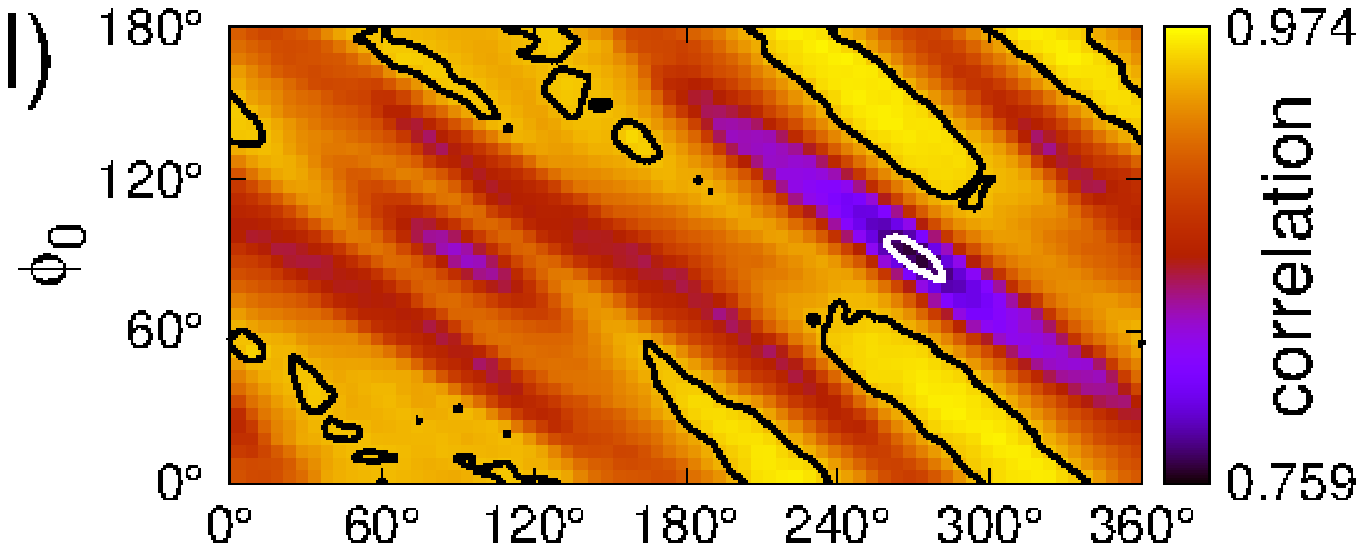}
\end{subfigure}

\begin{subfigure}[h]{0.42\textwidth}
\includegraphics[width=1.0\textwidth,angle=0]{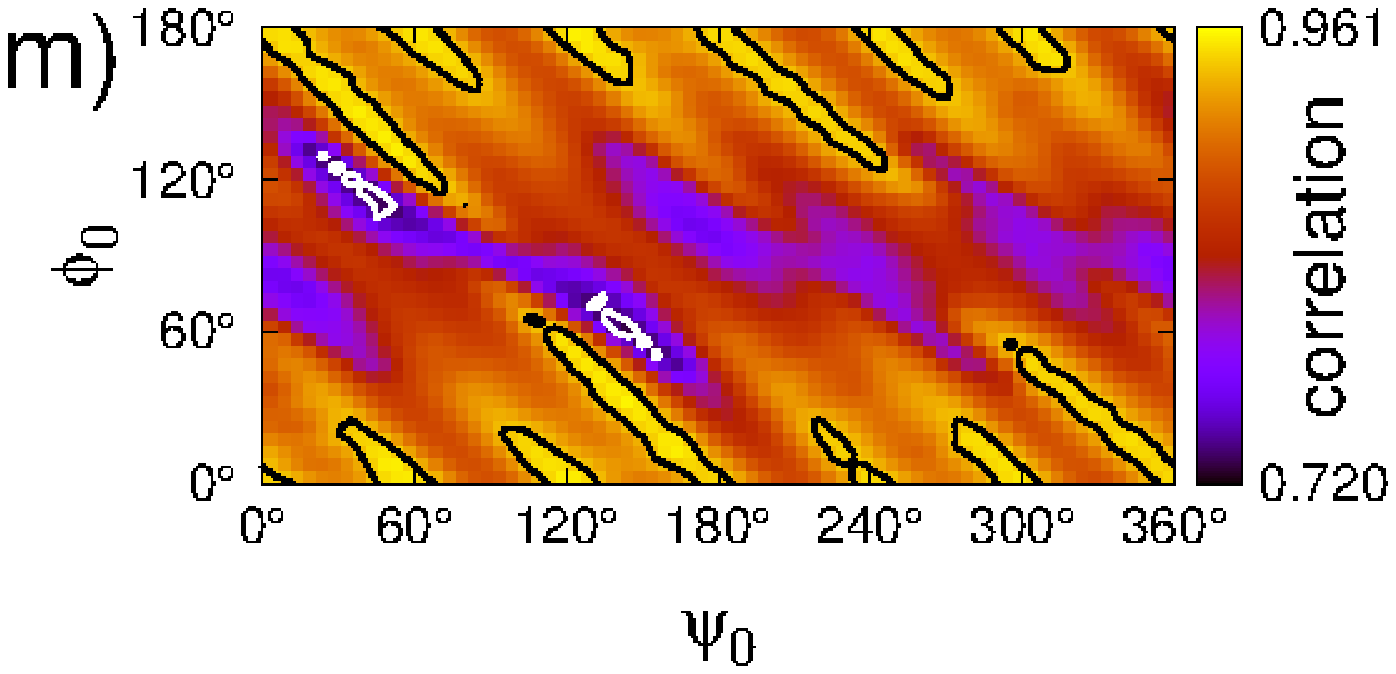}
\end{subfigure}
\begin{subfigure}[h]{0.42\textwidth}
\includegraphics[width=1.0\textwidth,angle=0]{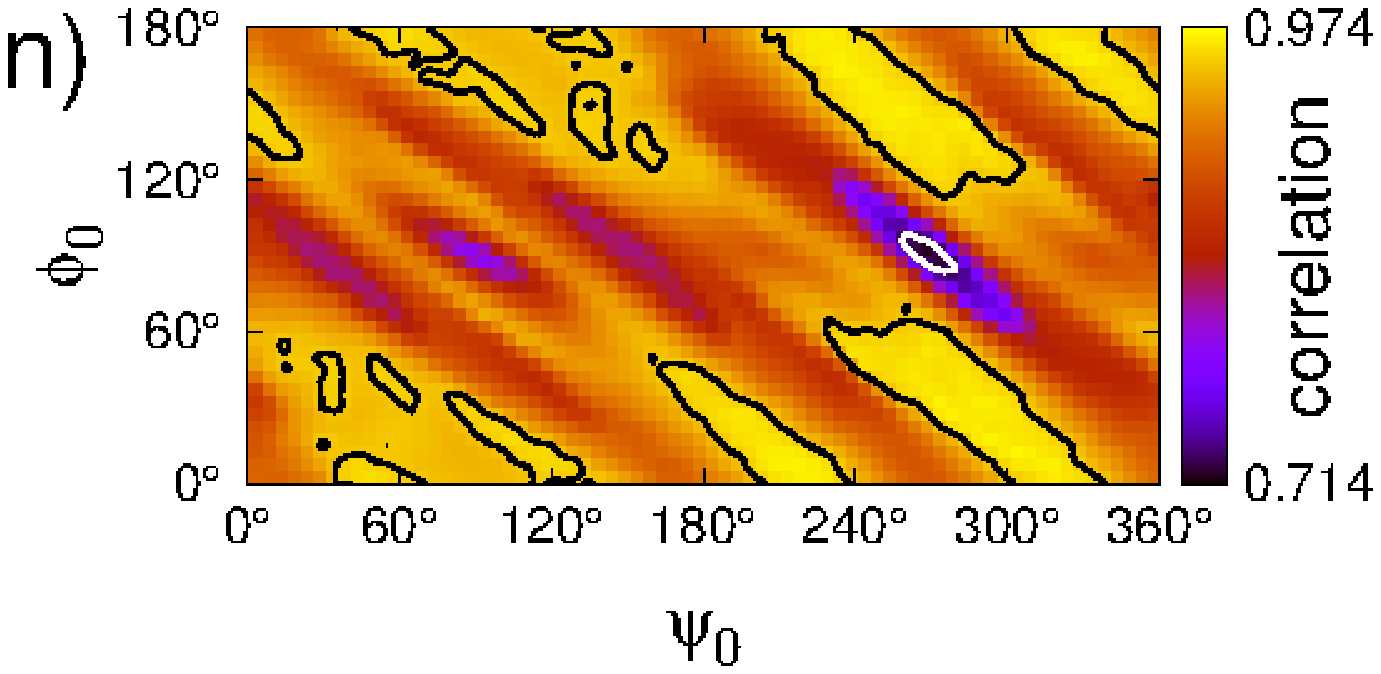}
\end{subfigure}
\caption{\label{Fig4} -1 V $\le V<0$ V negative bias range correlation analysis. Relative brightness correlation
distributions $r(\theta_0,\phi_0,\psi_0)$ for $\mathrm{W_{blunt}}$ tip [first column: a), c), e), g), i), k), m)] and
$\mathrm{W_{sharp}}$ tip [second column: b), d), f), h), j), l), n)] for the following fixed $\theta_0$ angles:
a)-b) $0^{\circ}$, c)-d) $5^{\circ}$, e)-f) $10^{\circ}$, g)-h) $15^{\circ}$, i)-j) $20^{\circ}$, k)-l) $25^{\circ}$,
m)-n) $30^{\circ}$.
Most (least) likely tip orientations in the experiment in the given bias interval correspond to bright (dark)
regions bounded by black (white) contours within 2\% relative to the maximum (minimum) correlation value in each subfigure
assuming the model tip apex geometry.
}
\end{figure*}

\begin{figure*}
\begin{subfigure}[h]{0.42\textwidth}
\includegraphics[width=1.0\textwidth,angle=0]{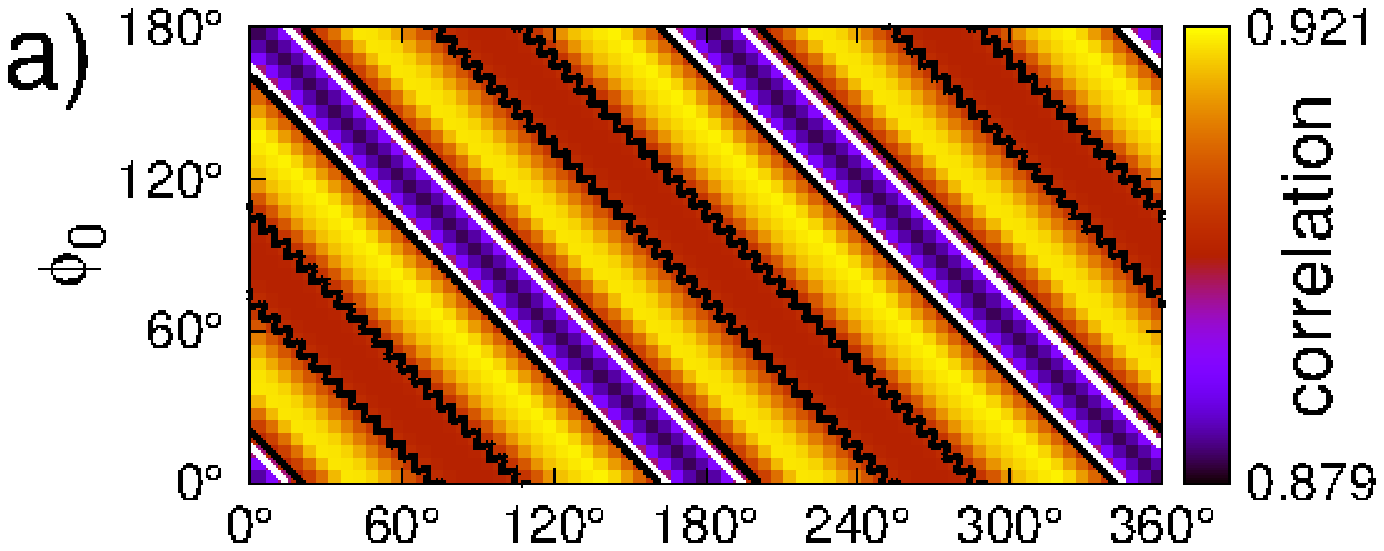}
\end{subfigure}
\begin{subfigure}[h]{0.42\textwidth}
\includegraphics[width=1.0\textwidth,angle=0]{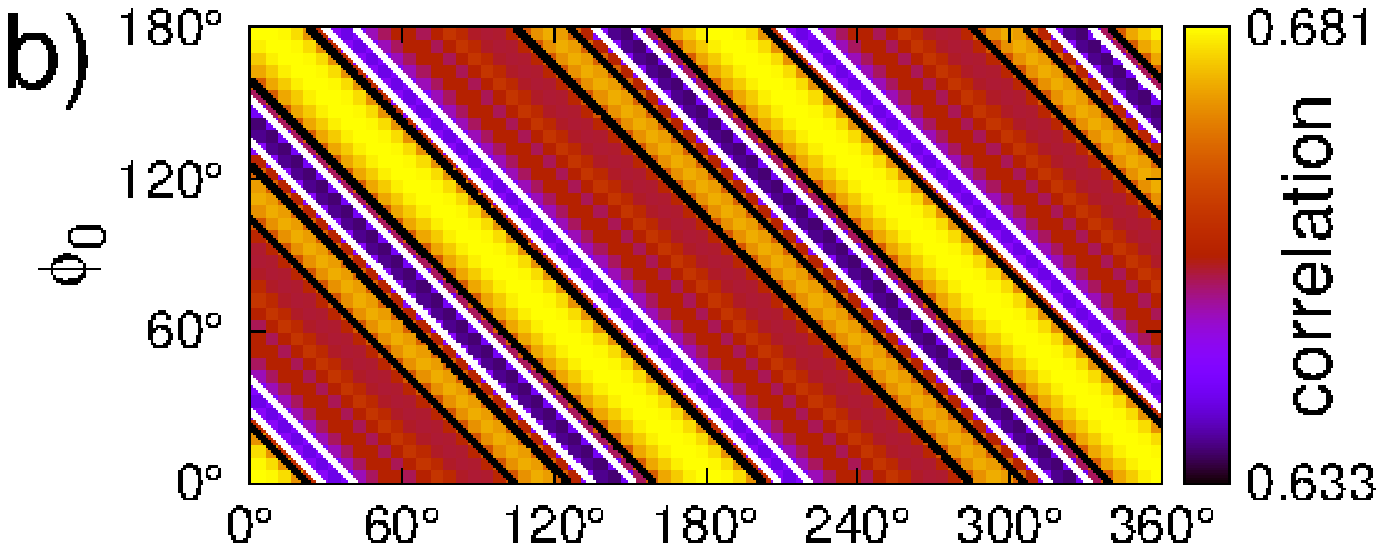}
\end{subfigure}

\begin{subfigure}[h]{0.42\textwidth}
\includegraphics[width=1.0\textwidth,angle=0]{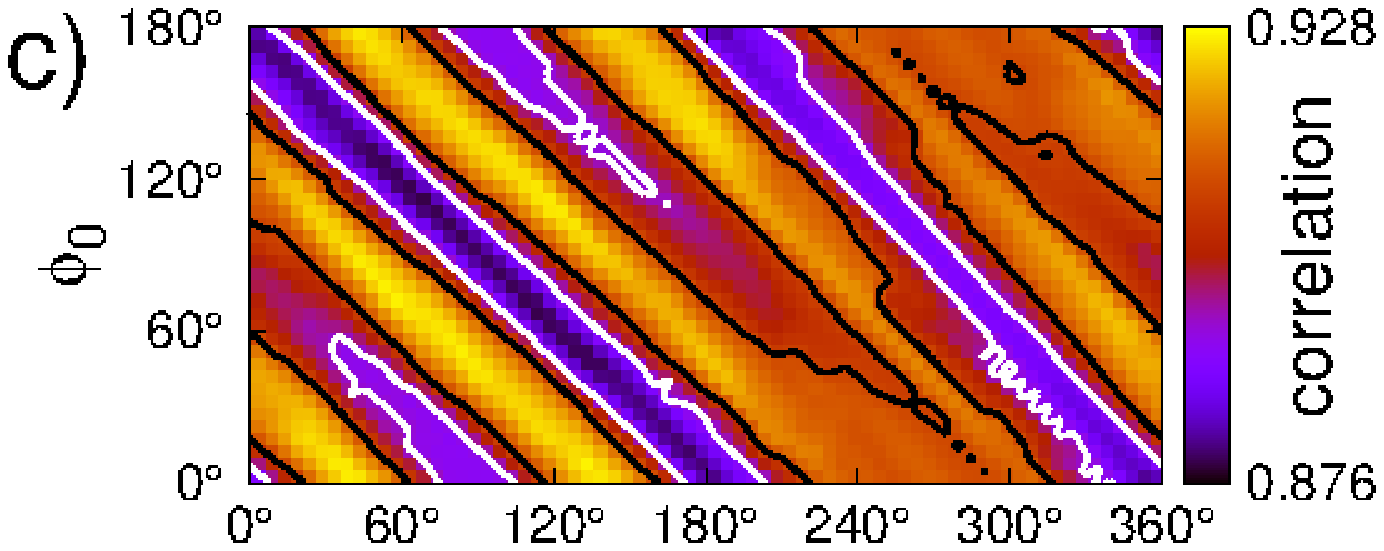}
\end{subfigure}
\begin{subfigure}[h]{0.42\textwidth}
\includegraphics[width=1.0\textwidth,angle=0]{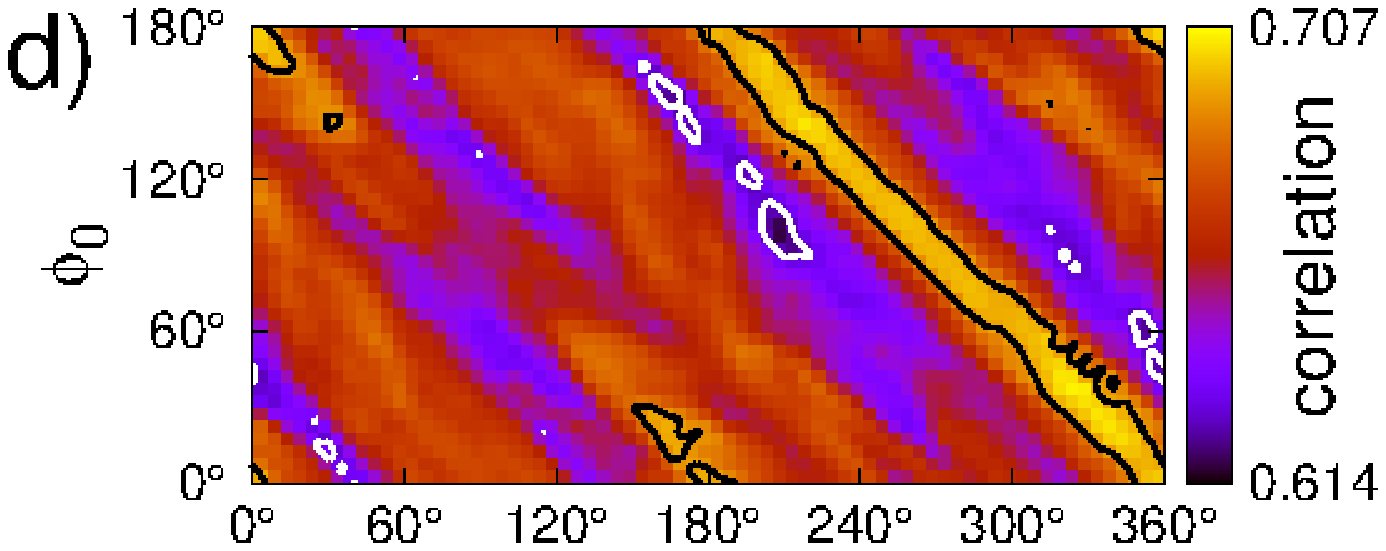}
\end{subfigure}

\begin{subfigure}[h]{0.42\textwidth}
\includegraphics[width=1.0\textwidth,angle=0]{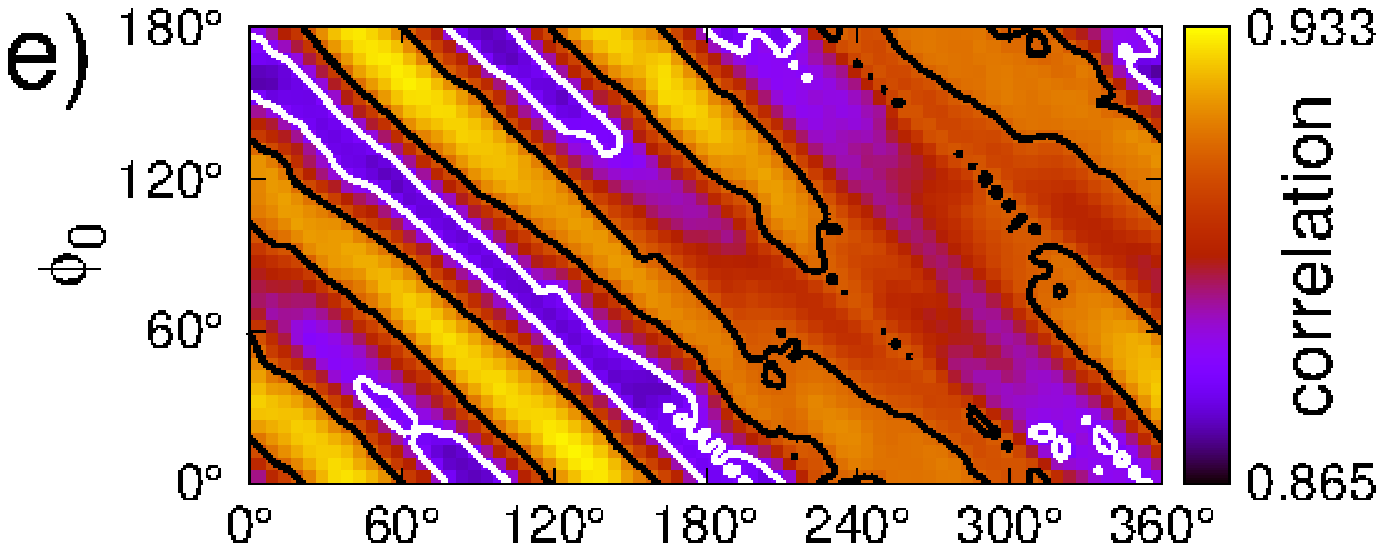}
\end{subfigure}
\begin{subfigure}[h]{0.42\textwidth}
\includegraphics[width=1.0\textwidth,angle=0]{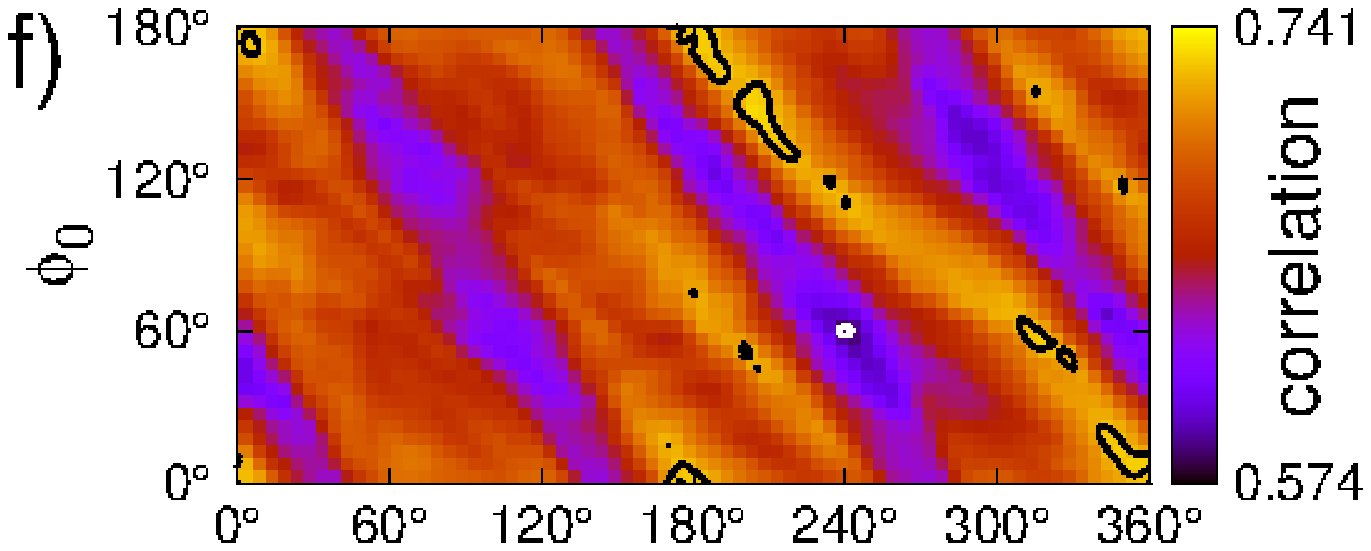}
\end{subfigure}

\begin{subfigure}[h]{0.42\textwidth}
\includegraphics[width=1.0\textwidth,angle=0]{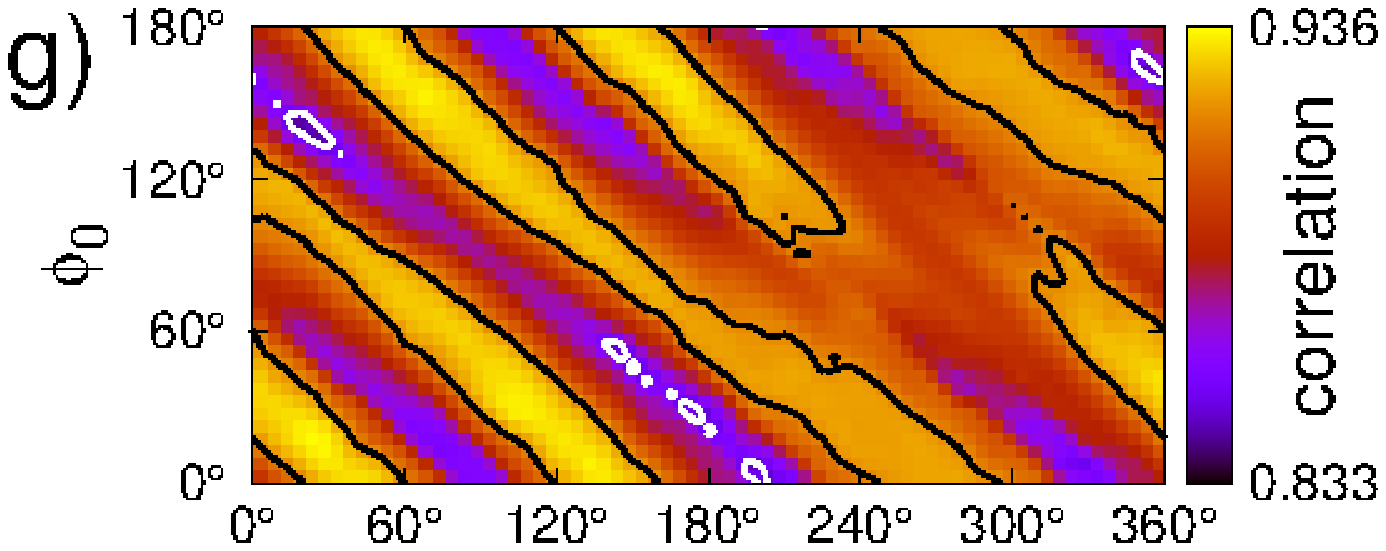}
\end{subfigure}
\begin{subfigure}[h]{0.42\textwidth}
\includegraphics[width=1.0\textwidth,angle=0]{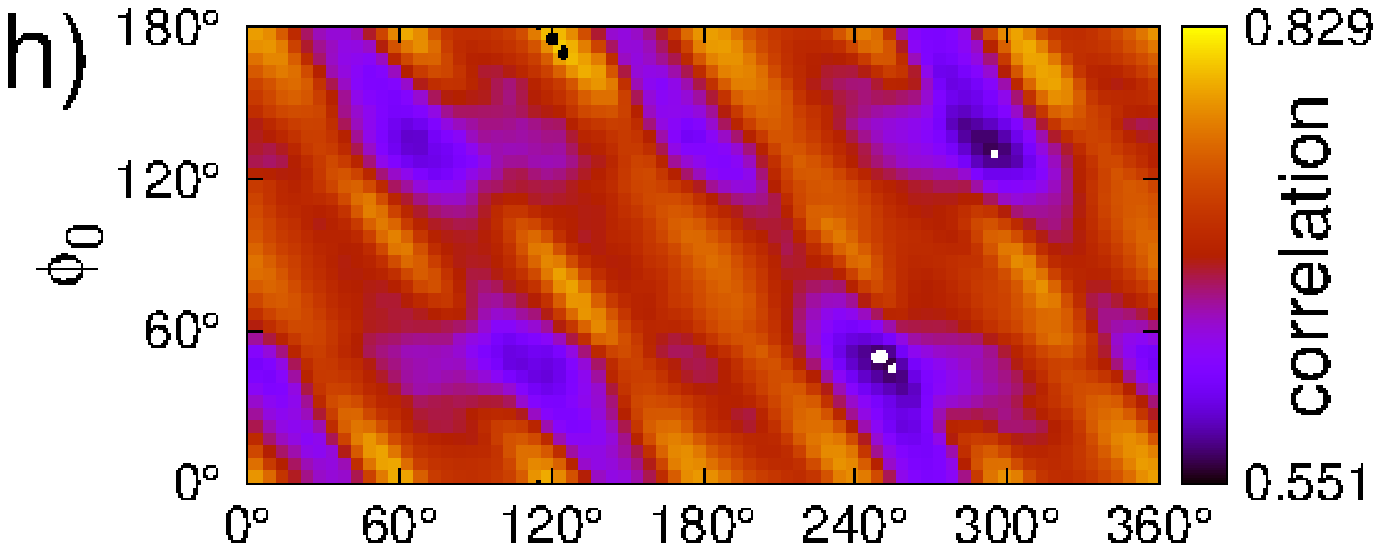}
\end{subfigure}

\begin{subfigure}[h]{0.42\textwidth}
\includegraphics[width=1.0\textwidth,angle=0]{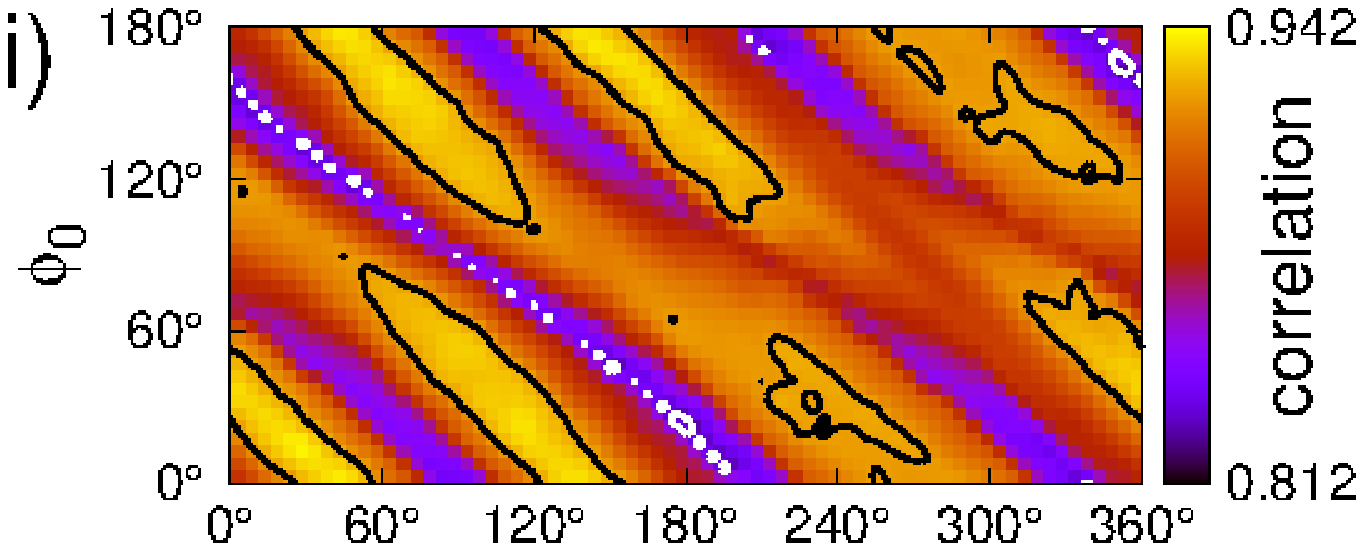}
\end{subfigure}
\begin{subfigure}[h]{0.42\textwidth}
\includegraphics[width=1.0\textwidth,angle=0]{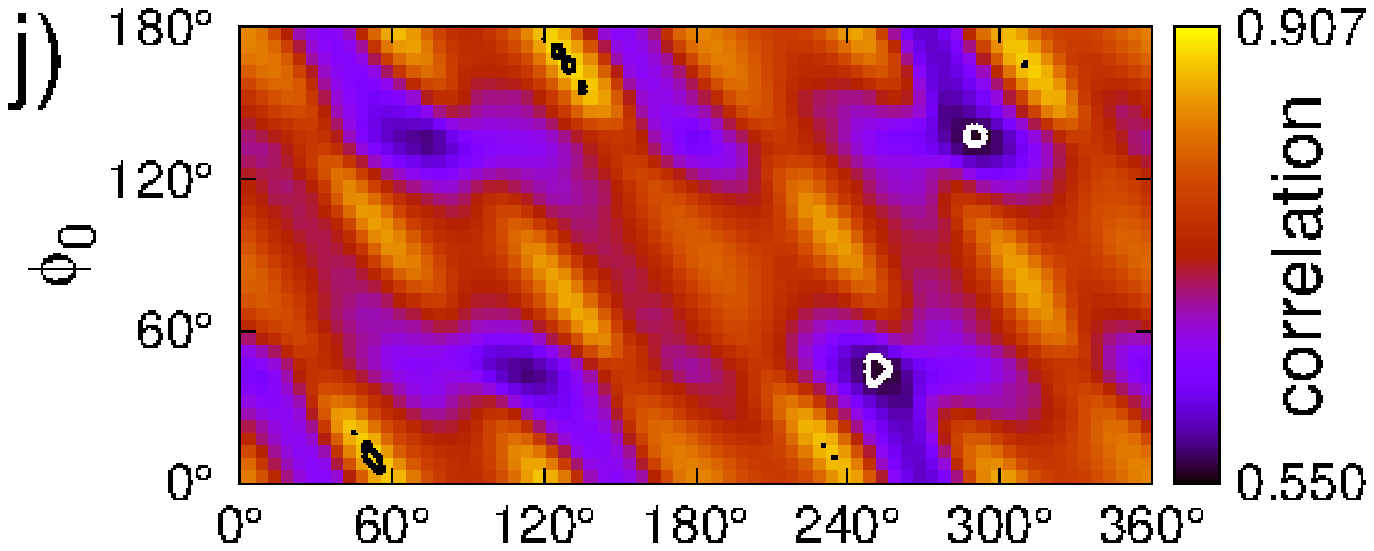}
\end{subfigure}

\begin{subfigure}[h]{0.42\textwidth}
\includegraphics[width=1.0\textwidth,angle=0]{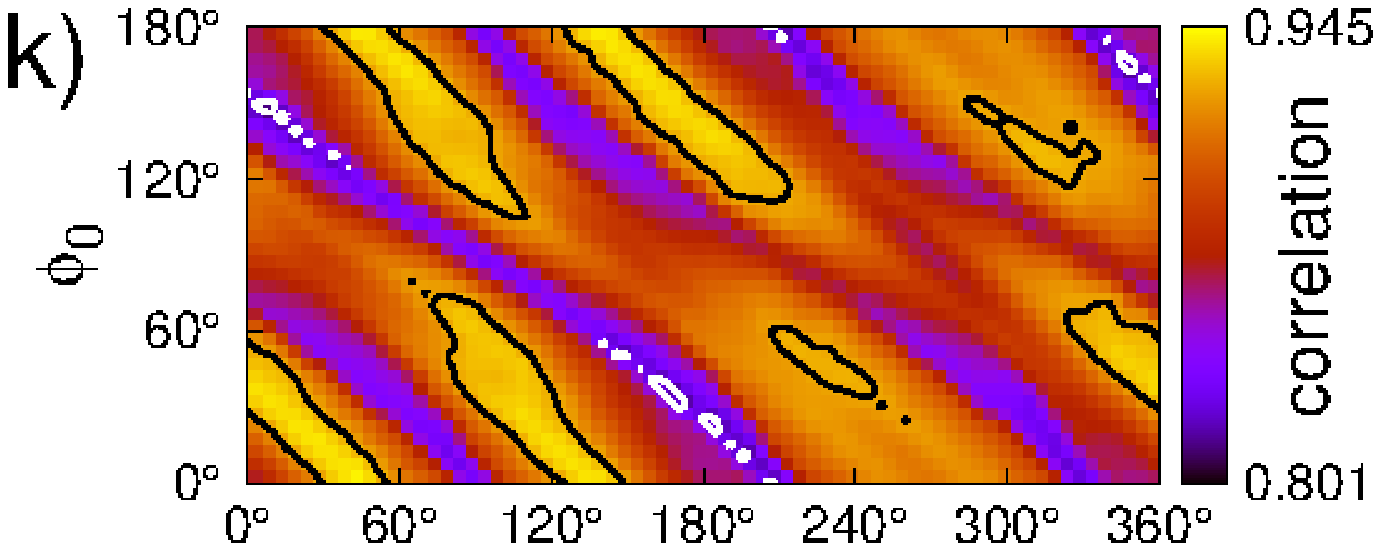}
\end{subfigure}
\begin{subfigure}[h]{0.42\textwidth}
\includegraphics[width=1.0\textwidth,angle=0]{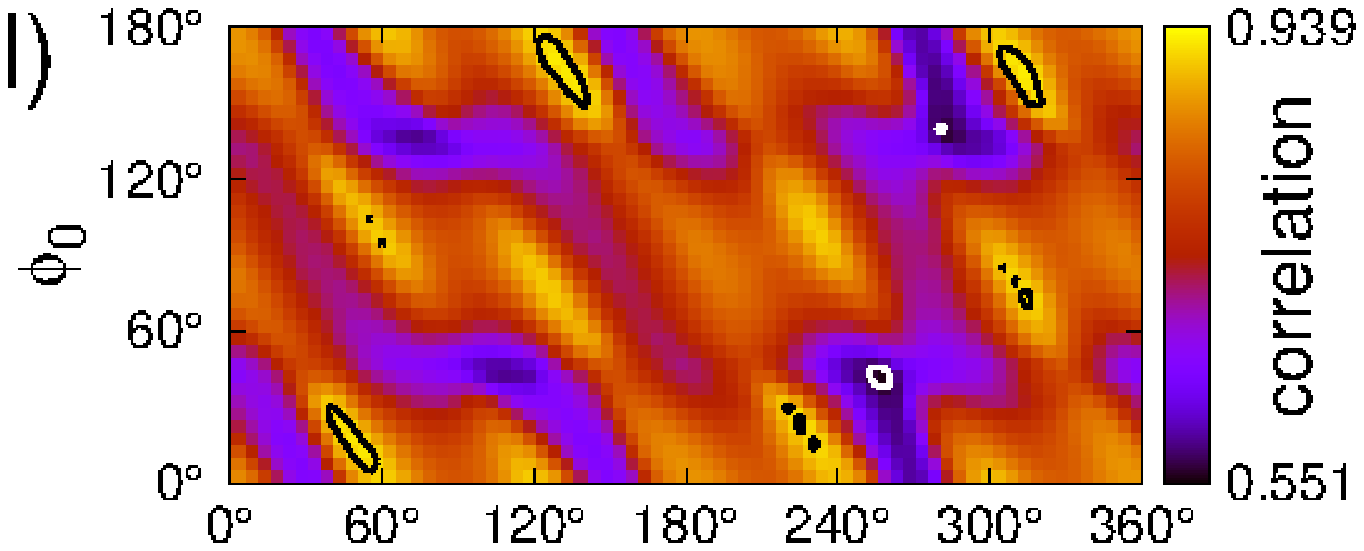}
\end{subfigure}

\begin{subfigure}[h]{0.42\textwidth}
\includegraphics[width=1.0\textwidth,angle=0]{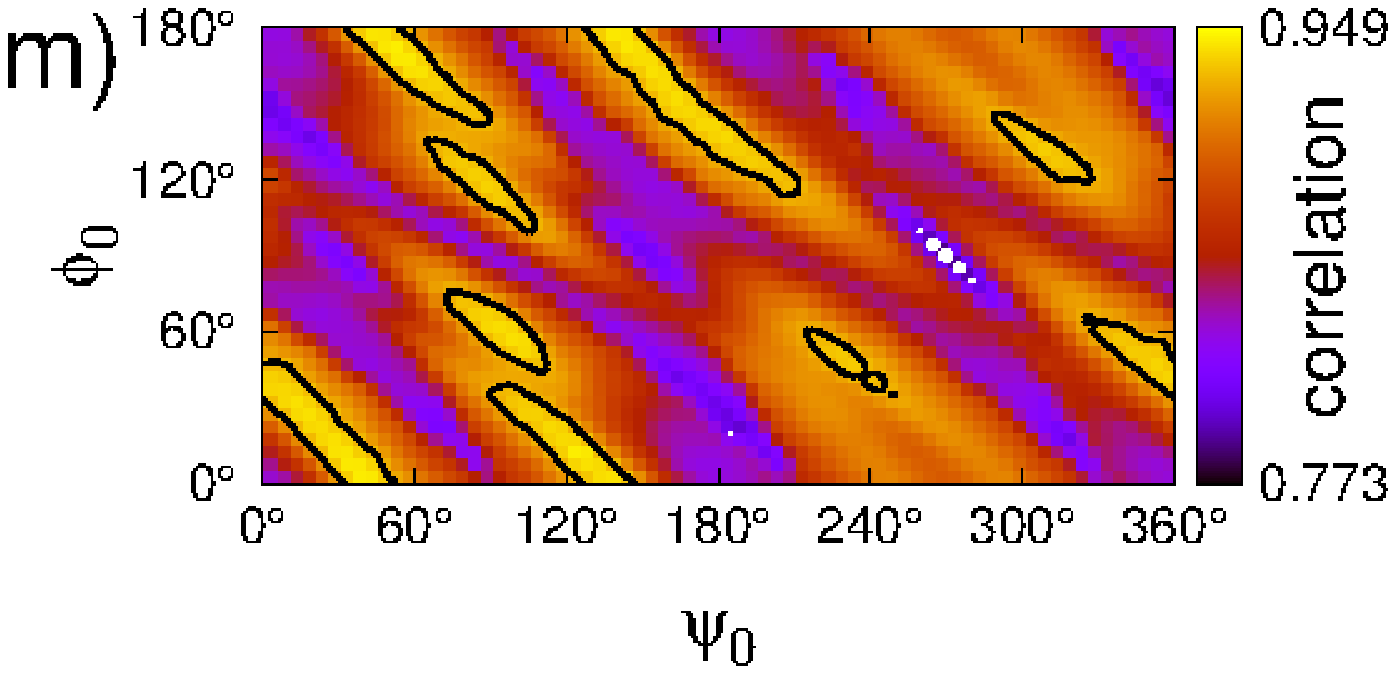}
\end{subfigure}
\begin{subfigure}[h]{0.42\textwidth}
\includegraphics[width=1.0\textwidth,angle=0]{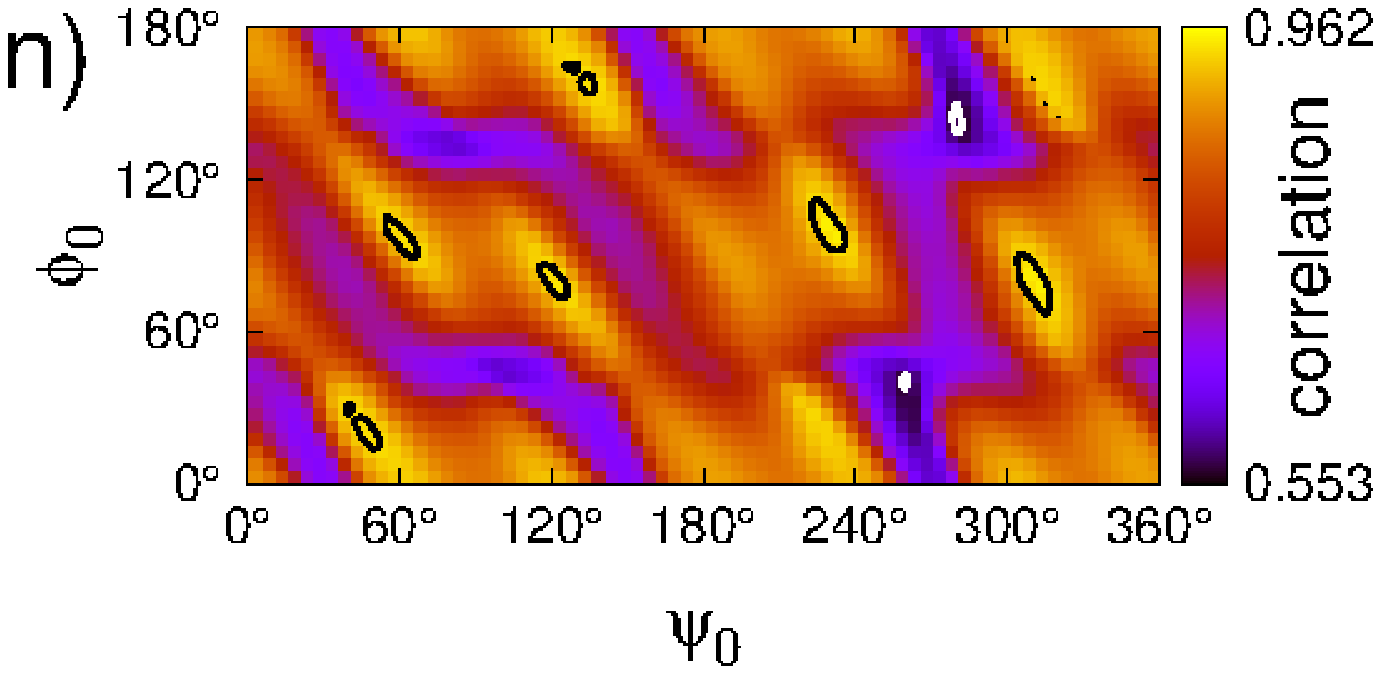}
\end{subfigure}
\caption{\label{Fig5} 0 V $<V\le$ 1 V positive bias range correlation analysis. Relative brightness correlation
distributions $r(\theta_0,\phi_0,\psi_0)$ for $\mathrm{W_{blunt}}$ tip [first column: a), c), e), g), i), k), m)] and
$\mathrm{W_{sharp}}$ tip [second column: b), d), f), h), j), l), n)] for the following fixed $\theta_0$ angles:
a)-b) $0^{\circ}$, c)-d) $5^{\circ}$, e)-f) $10^{\circ}$, g)-h) $15^{\circ}$, i)-j) $20^{\circ}$, k)-l) $25^{\circ}$,
m)-n) $30^{\circ}$.
Most (least) likely tip orientations in the experiment in the given bias interval correspond to bright (dark)
regions bounded by black (white) contours within 2\% relative to the maximum (minimum) correlation value in each subfigure
assuming the model tip apex geometry.
}
\end{figure*}

\begin{figure*}
\begin{subfigure}[h]{0.42\textwidth}
\includegraphics[width=1.0\textwidth,angle=0]{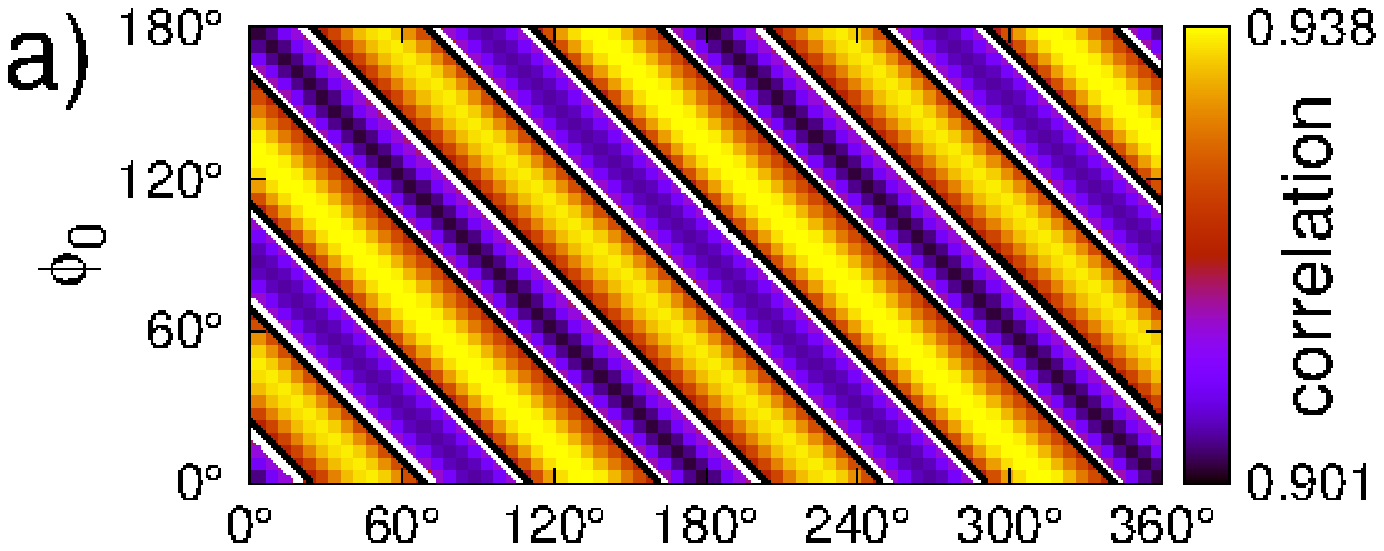}
\end{subfigure}
\begin{subfigure}[h]{0.42\textwidth}
\includegraphics[width=1.0\textwidth,angle=0]{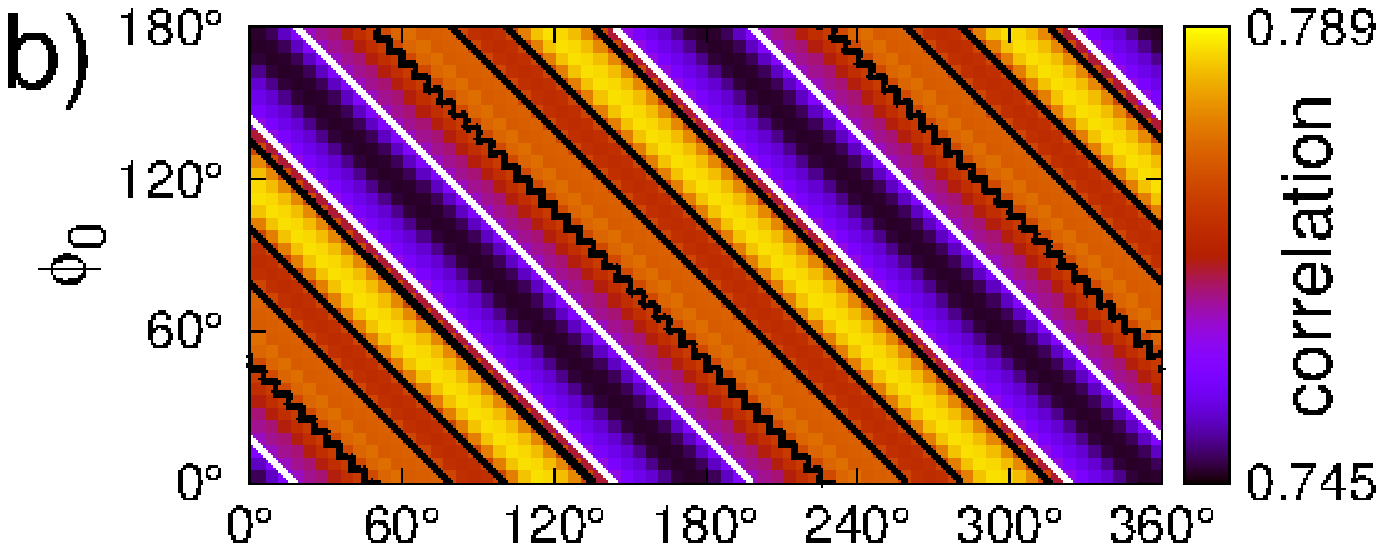}
\end{subfigure}

\begin{subfigure}[h]{0.42\textwidth}
\includegraphics[width=1.0\textwidth,angle=0]{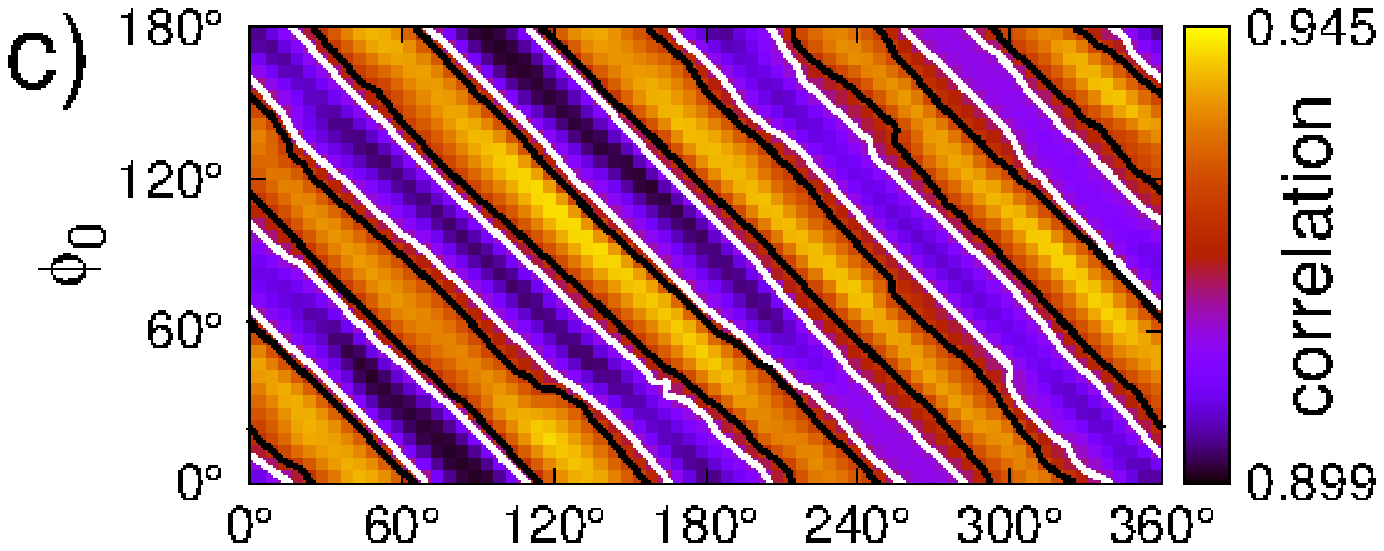}
\end{subfigure}
\begin{subfigure}[h]{0.42\textwidth}
\includegraphics[width=1.0\textwidth,angle=0]{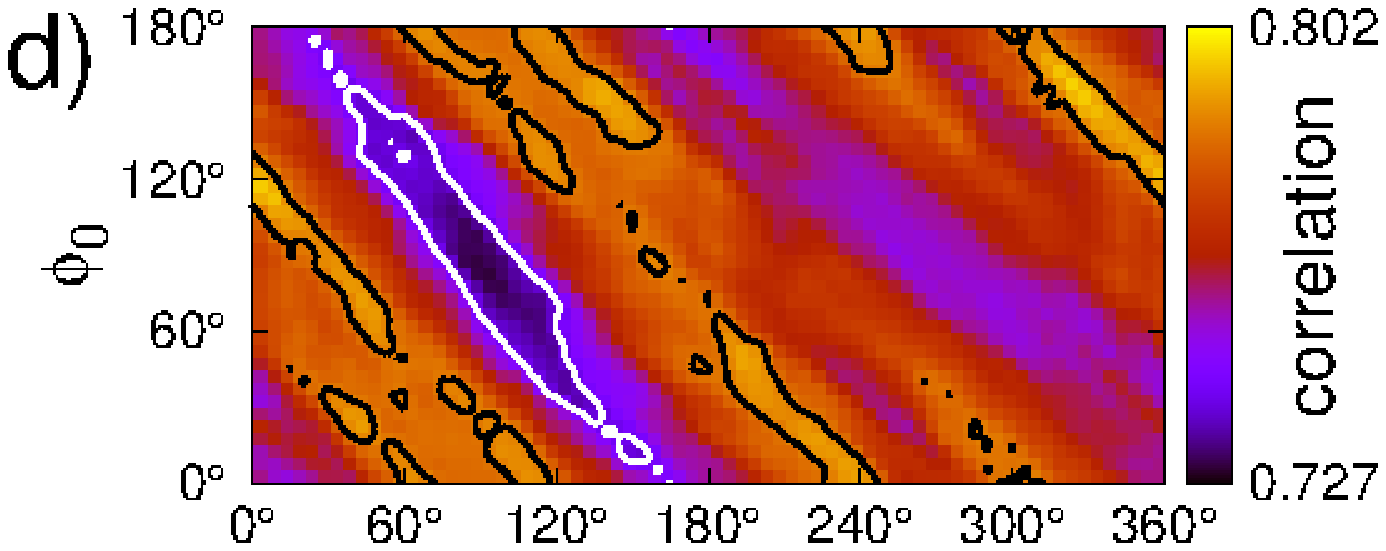}
\end{subfigure}

\begin{subfigure}[h]{0.42\textwidth}
\includegraphics[width=1.0\textwidth,angle=0]{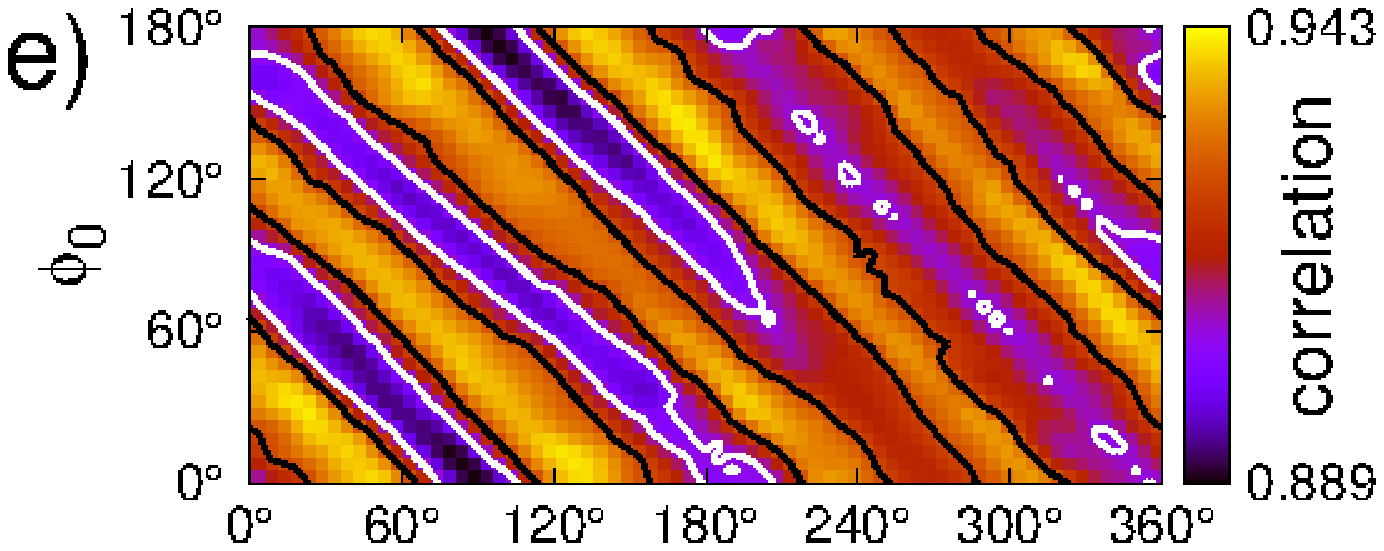}
\end{subfigure}
\begin{subfigure}[h]{0.42\textwidth}
\includegraphics[width=1.0\textwidth,angle=0]{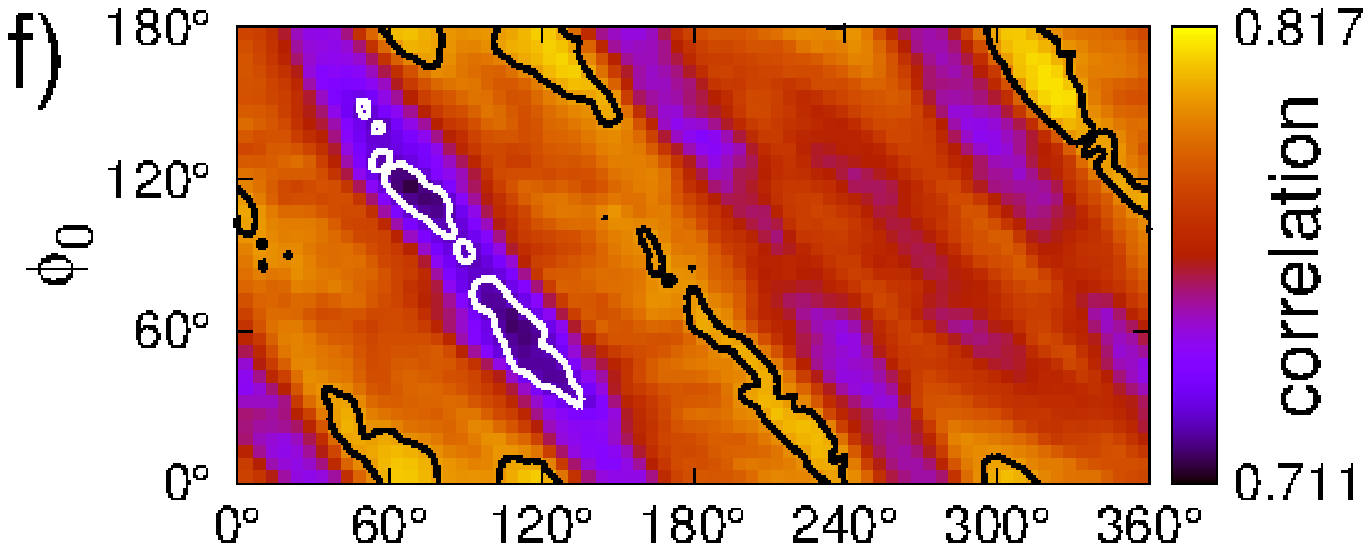}
\end{subfigure}

\begin{subfigure}[h]{0.42\textwidth}
\includegraphics[width=1.0\textwidth,angle=0]{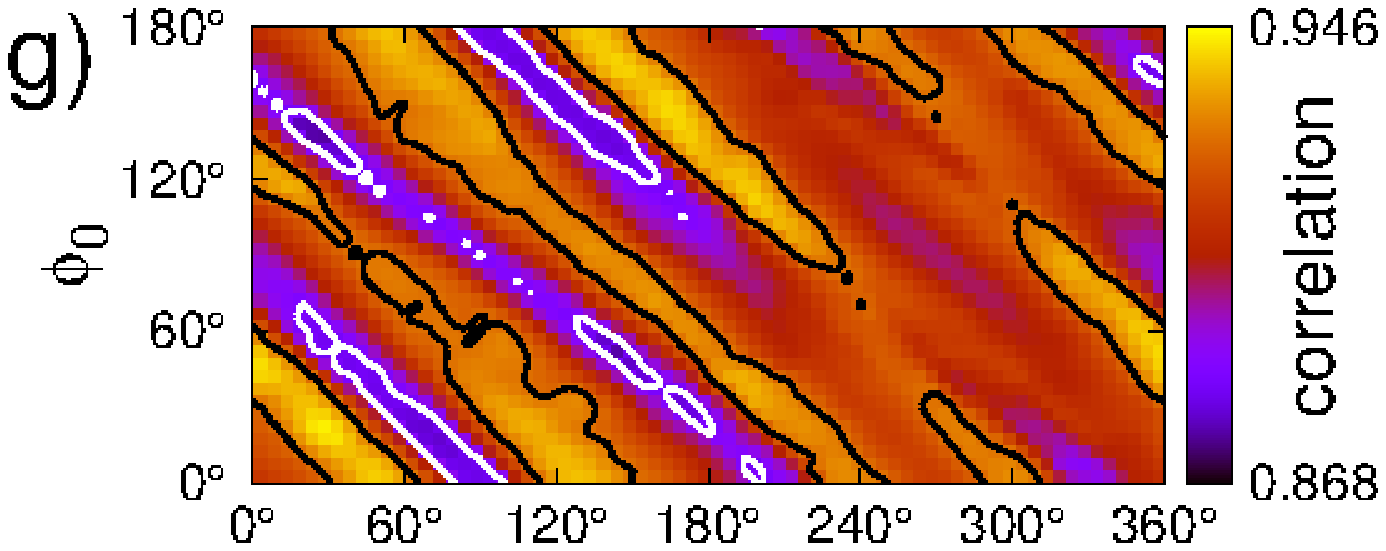}
\end{subfigure}
\begin{subfigure}[h]{0.42\textwidth}
\includegraphics[width=1.0\textwidth,angle=0]{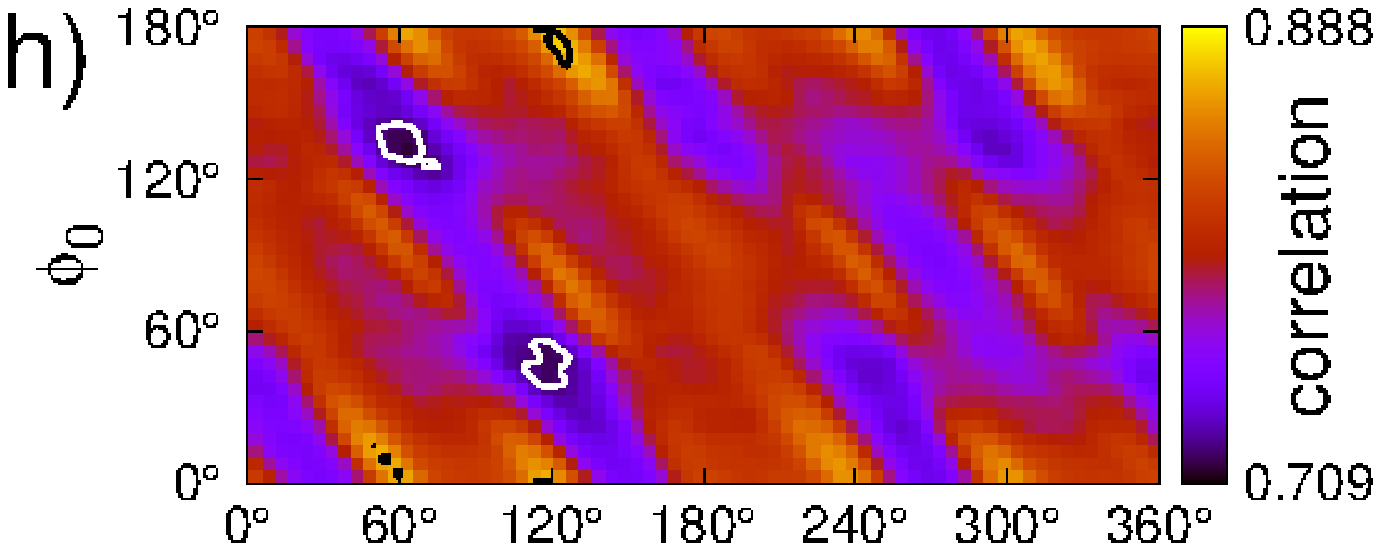}
\end{subfigure}

\begin{subfigure}[h]{0.42\textwidth}
\includegraphics[width=1.0\textwidth,angle=0]{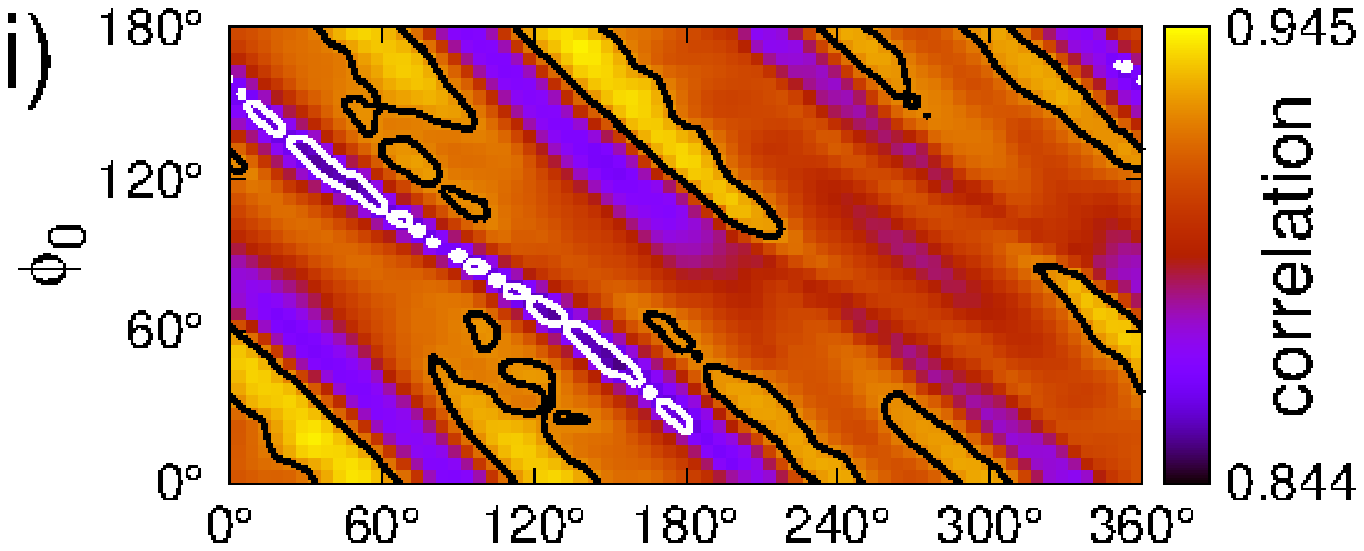}
\end{subfigure}
\begin{subfigure}[h]{0.42\textwidth}
\includegraphics[width=1.0\textwidth,angle=0]{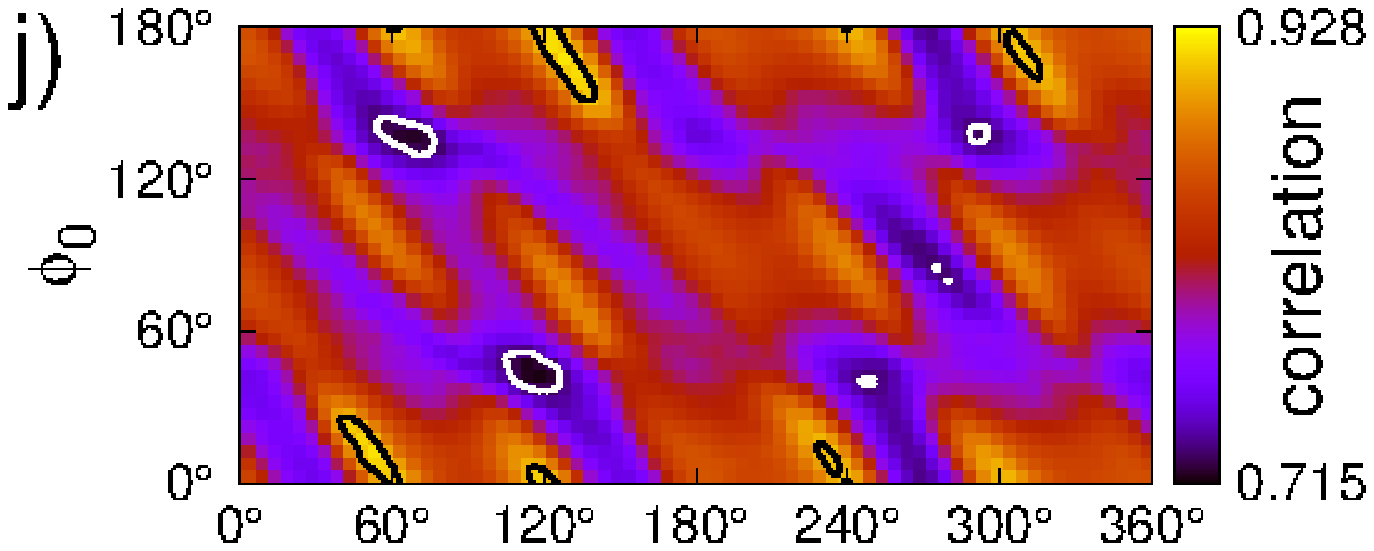}
\end{subfigure}

\begin{subfigure}[h]{0.42\textwidth}
\includegraphics[width=1.0\textwidth,angle=0]{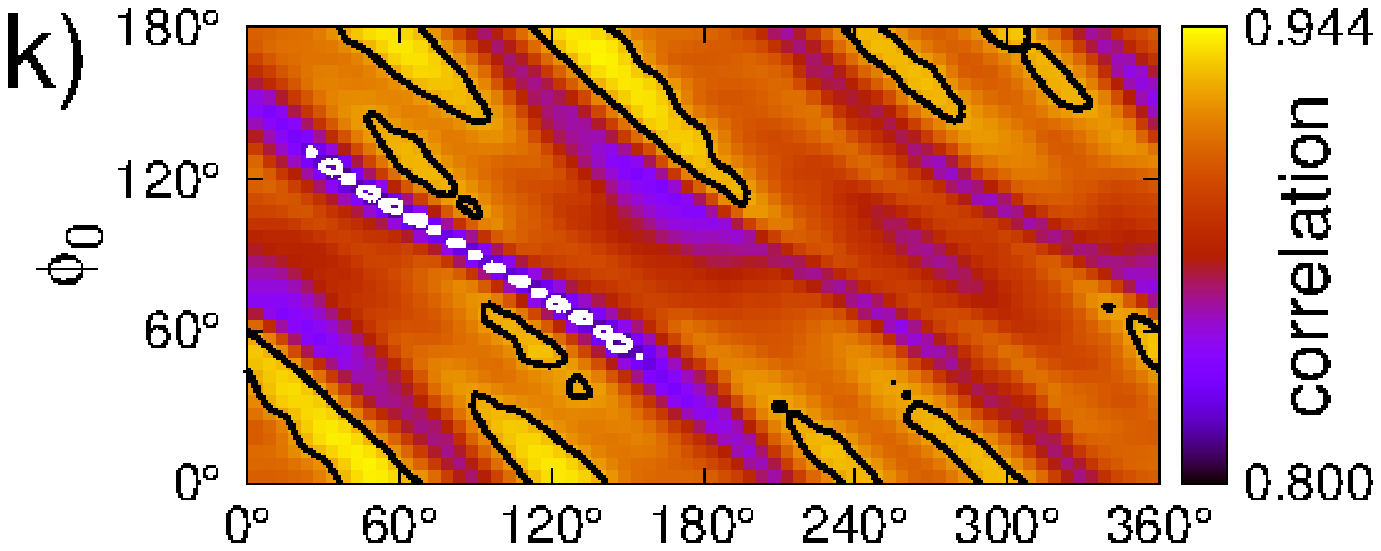}
\end{subfigure}
\begin{subfigure}[h]{0.42\textwidth}
\includegraphics[width=1.0\textwidth,angle=0]{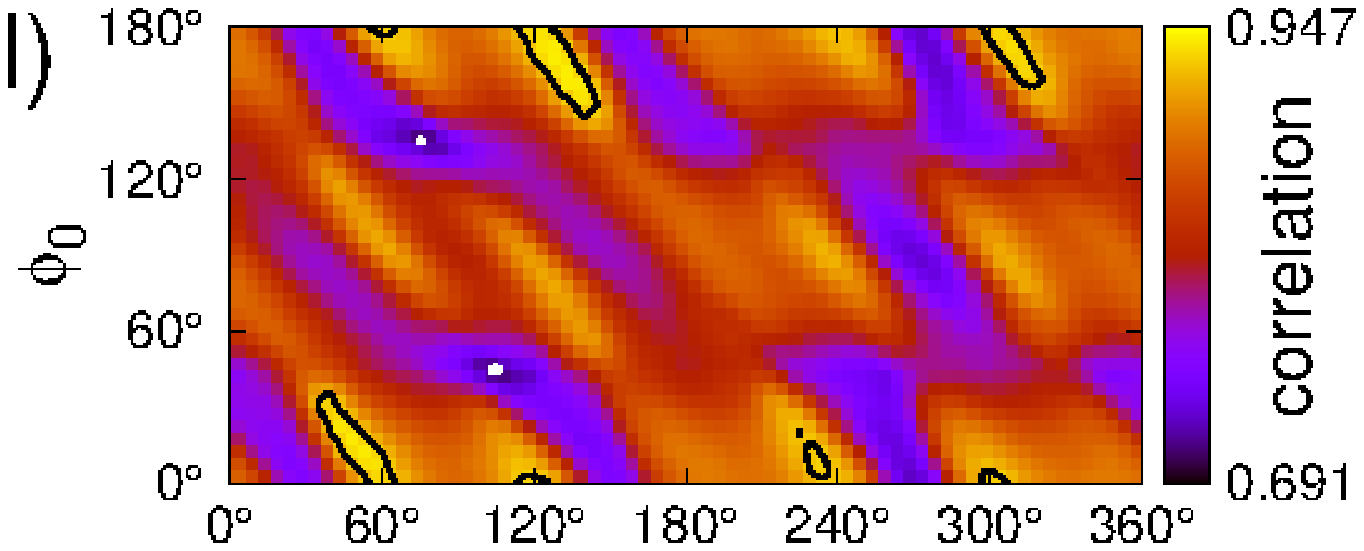}
\end{subfigure}

\begin{subfigure}[h]{0.42\textwidth}
\includegraphics[width=1.0\textwidth,angle=0]{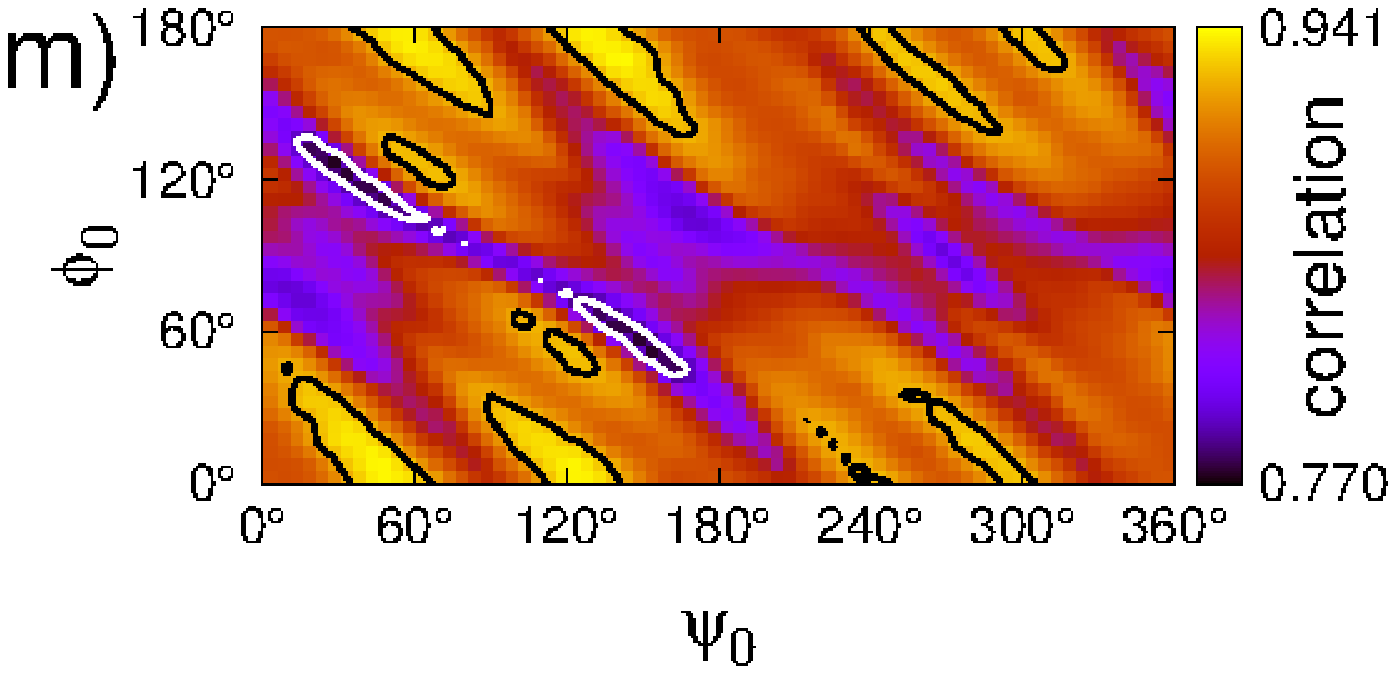}
\end{subfigure}
\begin{subfigure}[h]{0.42\textwidth}
\includegraphics[width=1.0\textwidth,angle=0]{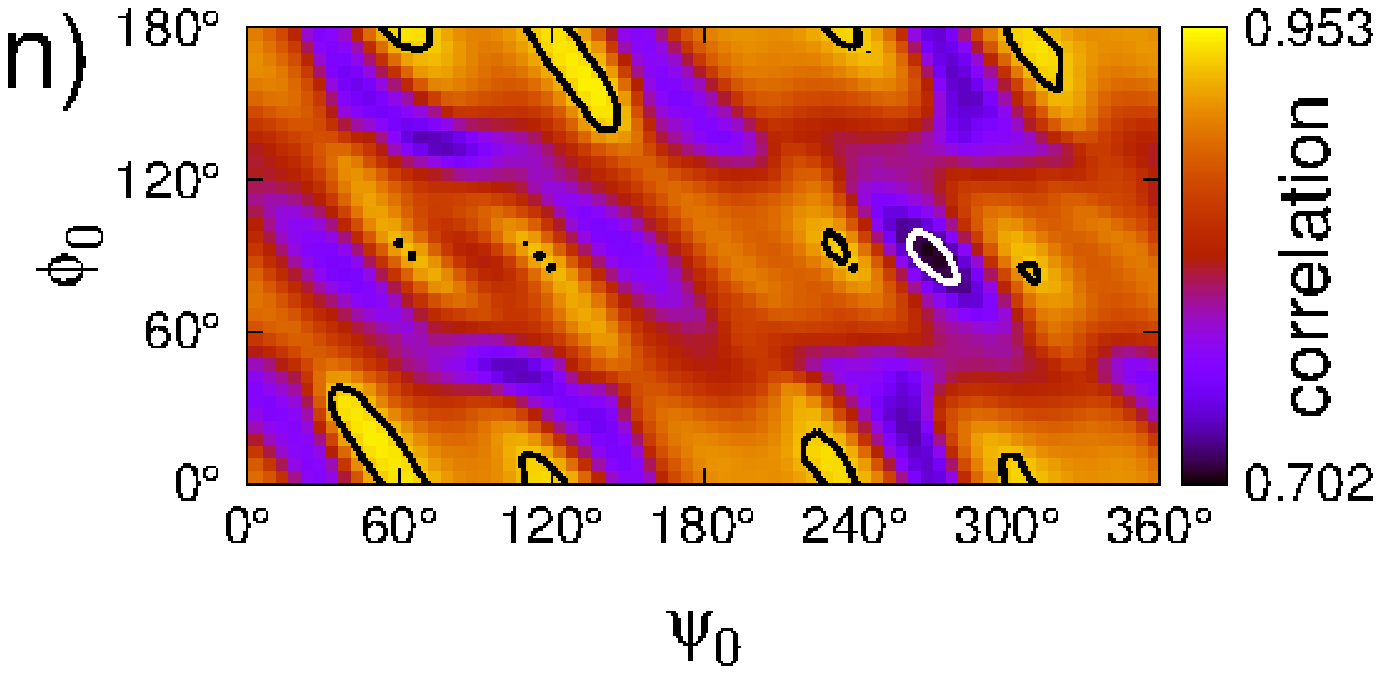}
\end{subfigure}
\caption{\label{Fig6} $|V|\le 1$ V full bias range correlation analysis. Relative brightness correlation
distributions $r(\theta_0,\phi_0,\psi_0)$ for $\mathrm{W_{blunt}}$ tip [first column: a), c), e), g), i), k), m)] and
$\mathrm{W_{sharp}}$ tip [second column: b), d), f), h), j), l), n)] for the following fixed $\theta_0$ angles:
a)-b) $0^{\circ}$, c)-d) $5^{\circ}$, e)-f) $10^{\circ}$, g)-h) $15^{\circ}$, i)-j) $20^{\circ}$, k)-l) $25^{\circ}$,
m)-n) $30^{\circ}$.
Most (least) likely tip orientations in the experiment in the given bias interval correspond to bright (dark)
regions bounded by black (white) contours within 2\% relative to the maximum (minimum) correlation value in each subfigure
assuming the model tip apex geometry.
}
\end{figure*}

Figs.\ \ref{Fig4}, \ref{Fig5} and \ref{Fig6} show the calculated relative brightness correlation maps
for the two considered tungsten tip models in the negative, positive
and full bias voltage range, respectively. $r(\phi_0,\psi_0)$ two-dimensional maps are shown as a function of $\theta_0$.
Note that $\theta_0=0^{\circ}$ corresponds to the same $z$-axis of the surface and the tip, and in this case $\phi_0$ and
$\psi_0$ denote the same type of rotations around the common $z$-axis. As a result, we obtain striped $r(\phi_0,\psi_0)$
correlation maps for $\theta_0=0^{\circ}$ [panels a) and b)]. For $\theta_0>0^{\circ}$ these maps quickly change to show
more complicated correlation distributions [panels c)-n)].
Most importantly, Figs.\ \ref{Fig4}, \ref{Fig5} and \ref{Fig6} show the most (least) likely tip orientations
$(\theta_0,\phi_0,\psi_0)$ in the experiment in the given bias interval corresponding to bright (dark) regions bounded by
black (white) contours within 2\% relative to the maximum (minimum) correlation value for each $\theta_0$ assuming the model
tip apex geometry. Overall, we find that the regions close to the maximal and minimal correlations can be differently affected
by the bias range considered for the mapping for different tip apex geometries. These results emphasize the importance of a
large experimental dataset for reliable application of the proposed procedure.
Considering the favorable and unfavorable orientations for the given tip models, we find that the $(\phi_0,\psi_0)$ positions of
the indicated regions close to the maximum and minimum correlations in the $r(\phi_0,\psi_0)$ maps are fairly stable
with respect to the change of $\theta_0$. This means that the specific $(\phi_0,\psi_0)$ Euler angles are representative for the
likely (bright regions) and unlikely (dark regions) tip orientations in the STM experiment, irrespective of $\theta_0$.
Based on our results, we find that the favored tip-sample relative orientations are far from being symmetric.

We introduce the area ratios as the number of tip orientations
(area) within the denoted regions in Figs.\ \ref{Fig4}, \ref{Fig5} and \ref{Fig6} divided by the area of the $r(\phi_0,\psi_0)$
maps ($36\times 72$). These area ratios at fixed $\theta_0$ can be interpreted as the likelihood of favorable or unfavorable tip
orientations in the experiment assuming the considered tip geometry in the given bias range. The area ratios alone, however,
are not sufficient to identify the most or least likely tip orientations in the experiment since the maximum and minimum
correlation values vary considerably depending on $\theta_0$.

\begin{figure*}
\includegraphics[width=1.00\textwidth,angle=0]{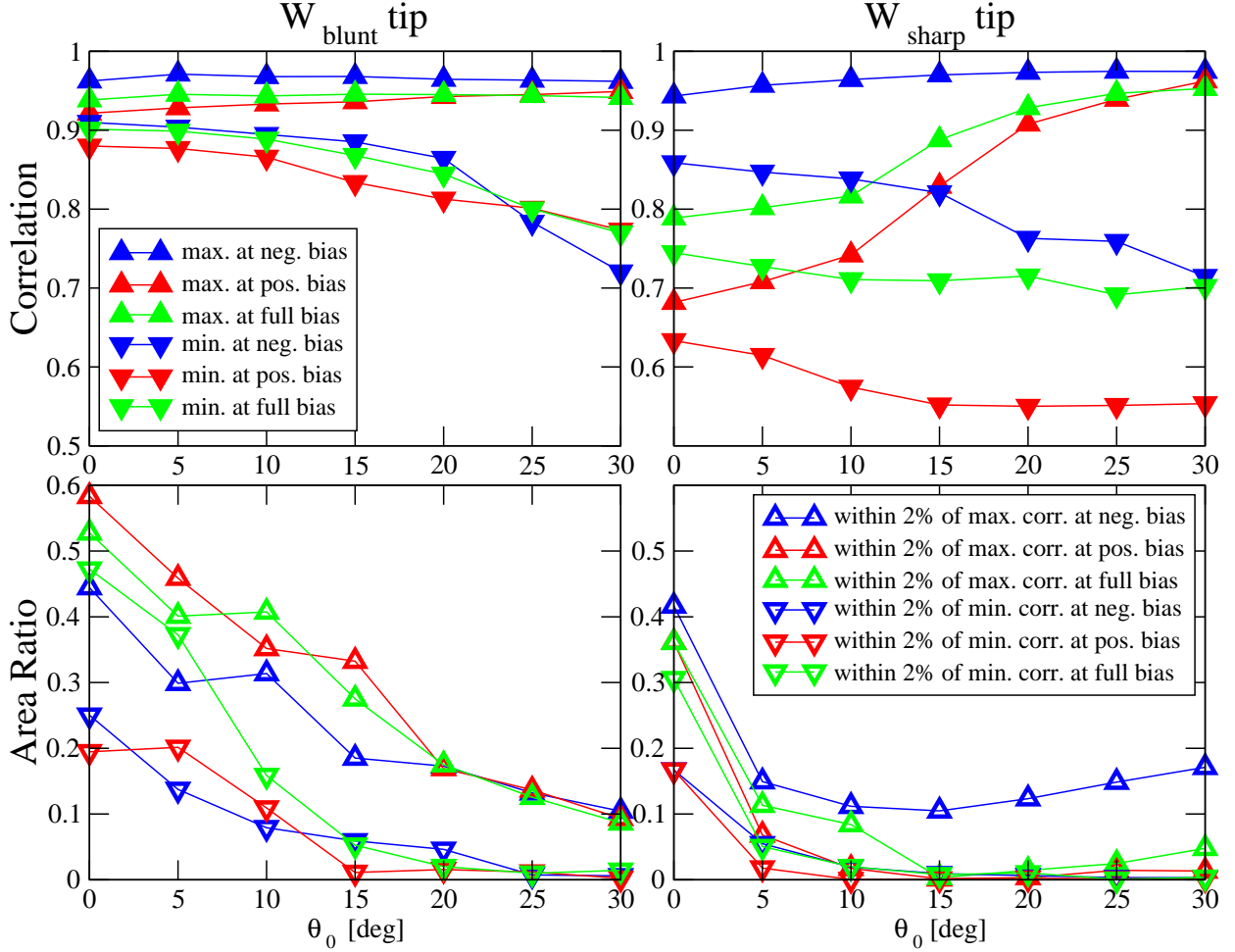}
\caption{\label{Fig7} Analysis of the correlation maps in Fig.\ \ref{Fig4} (at negative bias),
Fig.\ \ref{Fig5} (at positive bias) and Fig.\ \ref{Fig6} (at full bias) in the $|V|\le 1$ V bias range.
Top row: The evolution of the maximum and minimum correlation value in the $r(\theta_0,\phi_0,\psi_0)$ maps with $\theta_0$.
Bottom row: The $\theta_0$-evolution of the area within 2\% relative to the maximum and minimum correlation values
(respectively bounded by the black and white contours in Figs.\ \ref{Fig4}, \ref{Fig5} and \ref{Fig6}) in relation to the
area of the $r(\phi_0,\psi_0)$ map ($36\times 72$). These area ratios at fixed $\theta_0$ can be interpreted as the likelihood of
favorable or unfavorable tip orientations in the experiment assuming the considered tip geometry.
Left and right parts respectively correspond to data obtained by $\mathrm{W_{blunt}}$ and $\mathrm{W_{sharp}}$ tip models.
}
\end{figure*}

To further analyze the correlation maps in Figs.\ \ref{Fig4}, \ref{Fig5} and \ref{Fig6}, the evolutions of the maximum and minimum
correlation values and the calculated area ratios with $\theta_0$ are reported in Fig.\ \ref{Fig7}.
This figure also allows comparison between the different bias voltage ranges and the two considered tip models.
We find that the maximum correlation is increasing and the minimum correlation is decreasing with increasing $\theta_0$ for all
bias voltage ranges. This results in a larger difference between the maximum and minimum correlations with increasing $\theta_0$.
It is interesting to note that the maximum correlation values are always larger than 0.9 for the $\mathrm{W_{blunt}}$ tip, whereas
this is true only in the negative bias range for the $\mathrm{W_{sharp}}$ tip. In the positive and full bias ranges the
maximum correlation above 0.9 is achieved for $\theta_0\ge 20^{\circ}$, i.e., for a much smaller number of considered tip
orientations. On the other hand, the minimum correlation values are always smaller for the $\mathrm{W_{sharp}}$ compared to the
$\mathrm{W_{blunt}}$ tip. These findings clearly suggest that the $\mathrm{W_{blunt}}$ tip is more likely to be present in the
experiment in an enhanced bias voltage range than the $\mathrm{W_{sharp}}$ tip.

In Fig.\ \ref{Fig7}, at negative bias voltages the two tips provide similar maximum correlation values as a function of $\theta_0$.
In such case the area ratios can be used to decide which tip is more likely in the experiment since the corresponding area ratios
are proportional to the number of tip orientations within the maximum correlation, and such larger area ratios favor a given tip.
We find that the area ratios are generally larger for the $\mathrm{W_{blunt}}$ compared to the $\mathrm{W_{sharp}}$ tip.
Area ratios close to the correlation maximum mean that more orientations can provide better correlation values for the
$\mathrm{W_{blunt}}$ than for the $\mathrm{W_{sharp}}$ tip. On the other hand, area ratios close to the correlation minimum
mean that more orientations provide correlations close to the minimum for the $\mathrm{W_{blunt}}$ compared to the
$\mathrm{W_{sharp}}$ tip. This is, however, not a problem in the present case since the minimum correlations are always larger for
the $\mathrm{W_{blunt}}$ compared to the $\mathrm{W_{sharp}}$ tip. Therefore, based on the number of favorable tip orientations,
we can also conclude that the blunt tungsten tip is indeed more likely in the experiment than the sharp tip
in the $|V|\le 1$ V bias voltage range.

In order to check the robustness of our results we performed the correlation analysis with simulated brightness profiles obtained
by taking the contributions of four extra next-neighbor atoms of the tip apex atom in the tunneling current calculations using the
3D-WKB method. We find that the correlation maps are quantitatively very similar to those obtained by the one-apex tip for
$\theta_0\le 20^{\circ}$. For larger $\theta_0$-tilting the emergence of multiple tip apices distorts the simulated brightness
profiles and consequently worsens the agreement with the experiment, manifesting as dramatically reduced correlation values
(down to 0.35 at $\theta_0=25^{\circ}$ and 0.13 at $\theta_0=30^{\circ}$) for particular $(\phi_0,\psi_0)$ ranges. Based on this,
we can conclude that our findings are robust for $\theta_0\le 20^{\circ}$, i.e. for a small tilting of the tip $z$-axis.

\begin{figure*}
\begin{subfigure}[h]{0.32\textwidth}
\includegraphics[width=1.0\textwidth,angle=0]{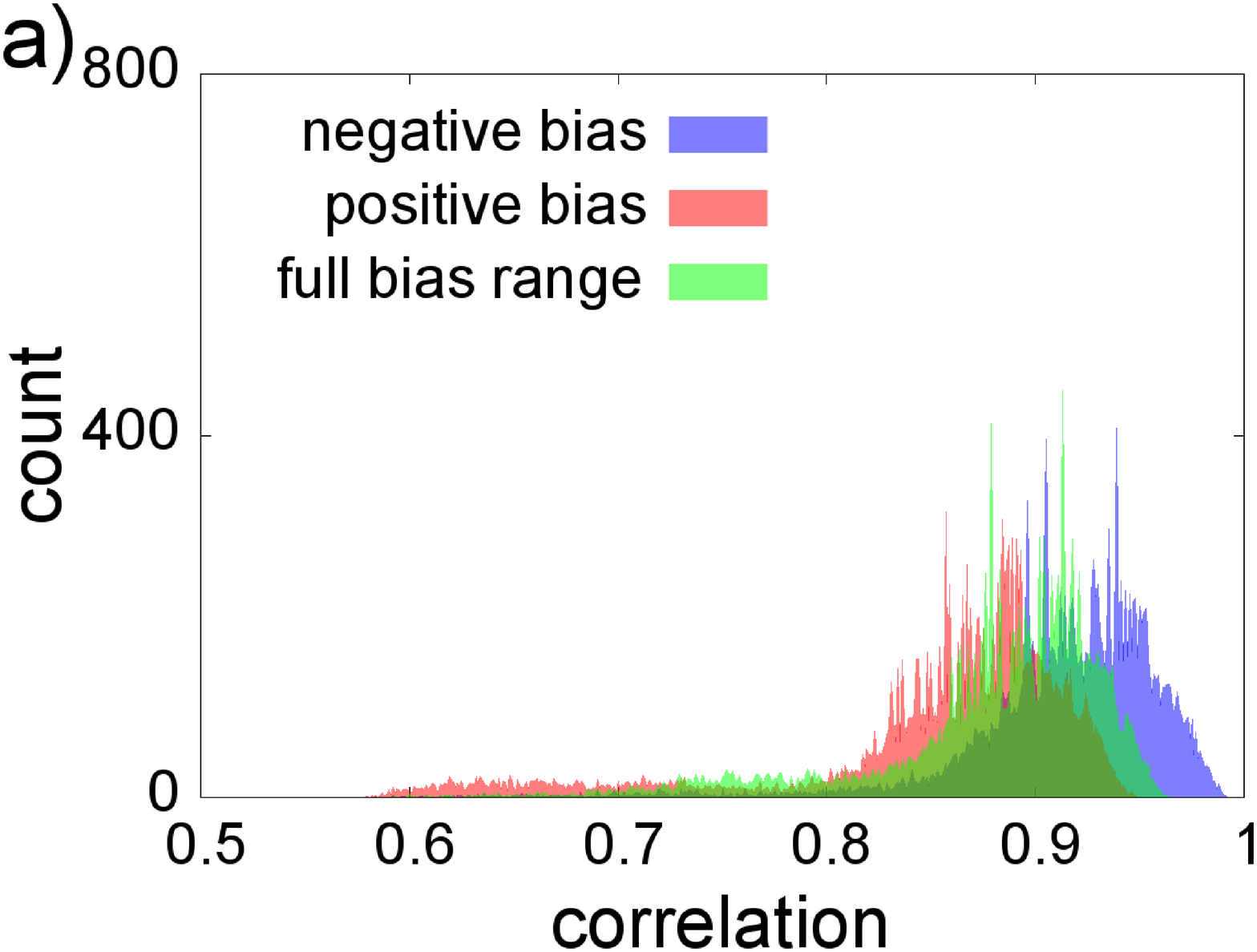}
\end{subfigure}
\begin{subfigure}[h]{0.32\textwidth}
\includegraphics[width=1.0\textwidth,angle=0]{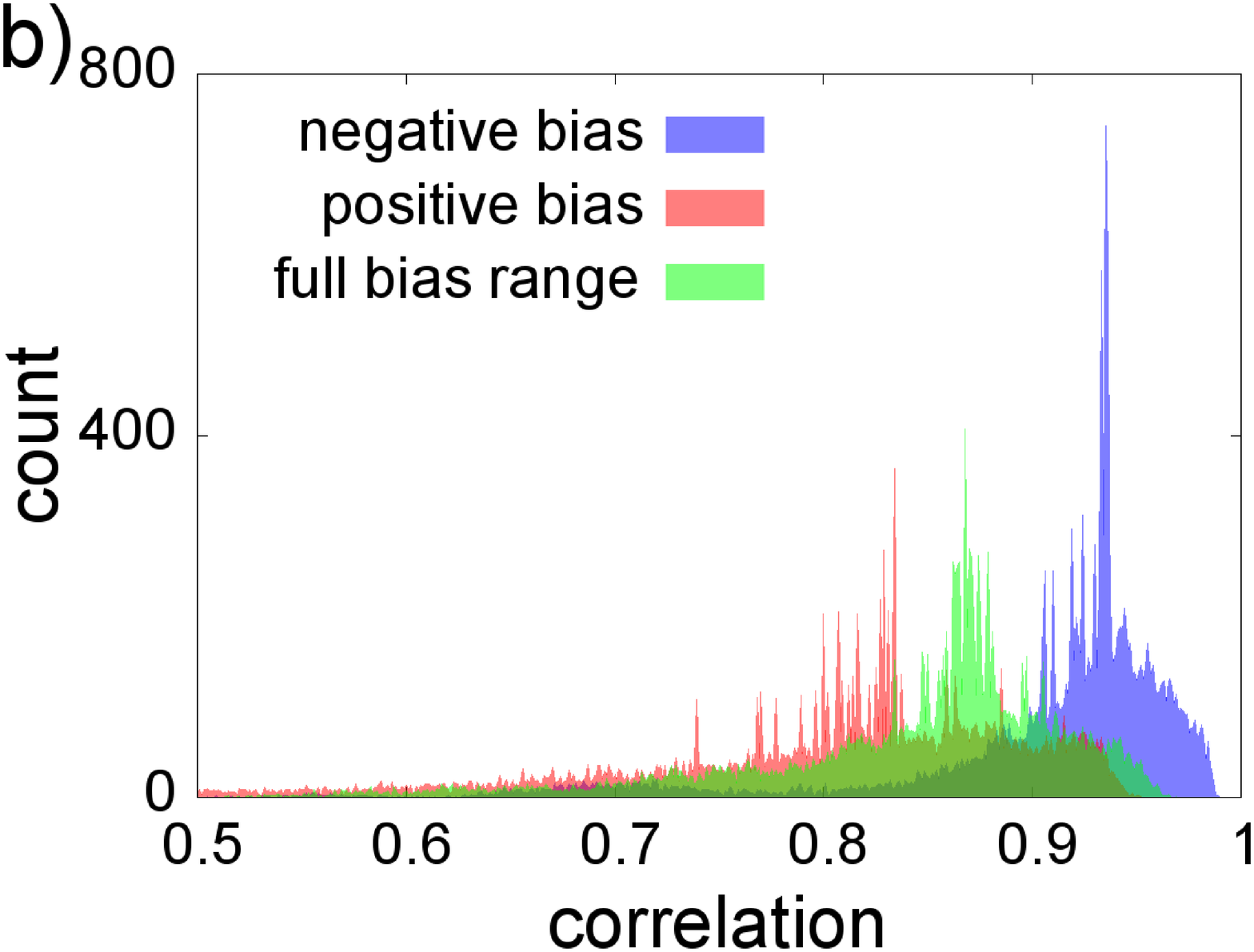}
\end{subfigure}
\begin{subfigure}[h]{0.32\textwidth}
\includegraphics[width=1.0\textwidth,angle=0]{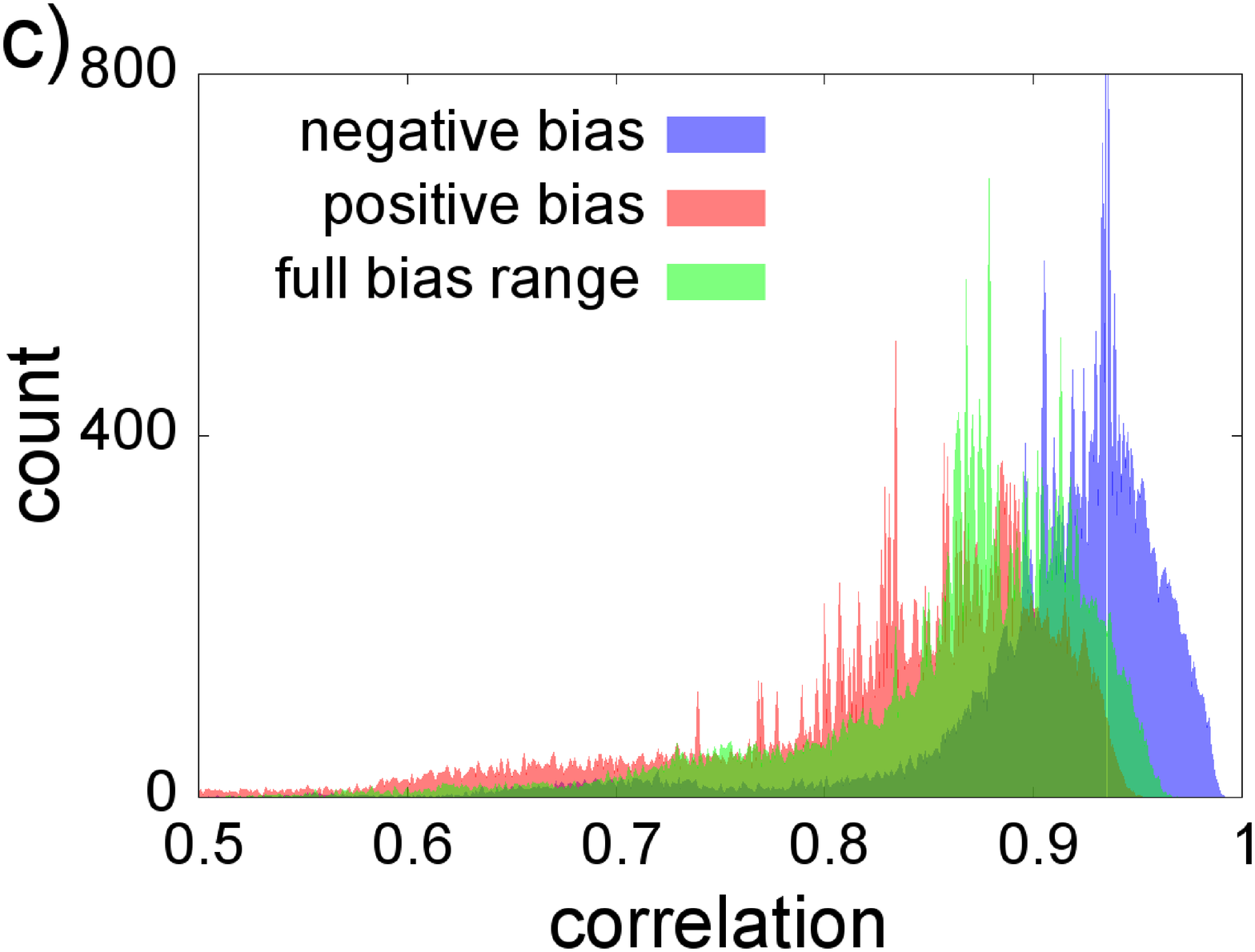}
\end{subfigure}
\caption{\label{Fig8} $|V|\le 0.3$ V relative brightness correlation histograms calculated by using 18144 tip orientations for:
a) $\mathrm{W_{blunt}}$ tip, b) $\mathrm{W_{sharp}}$ tip. Part c) reports the sum of the histograms in a) and b).
The correlation histograms for the negative, positive and full bias ranges are shown using Eq.(\ref{Eq_correlation}) in the
[0.5, 1] range with 0.001 resolution.
}
\end{figure*}

To investigate the effect of the bias voltage on the obtained results, we recalculated the correlation statistics in the
$|V|\le 0.3$ V bias voltage range that corresponds to the low bias regime used in typical STM imaging experiments of HOPG.
This analysis used redefined negative (-0.3 V $\le V<0$ V, $N_V=3$), positive (0 V $<V\le$ 0.3 V, $N_V=3$) and full
(-0.3 V $\le V\le$ 0.3 V, $N_V=6$) bias ranges. Fig.\ \ref{Fig8} shows the recalculated relative brightness correlation histograms
for the two considered tungsten tip models in 18144 tip orientations and the sum of the two histograms in Fig.\ \ref{Fig8}c).
We find qualitatively similar results as in the $|V|\le 1$ V bias range reported in Fig.\ \ref{Fig3}.
The main differences in Fig.\ \ref{Fig8} in comparison to Fig.\ \ref{Fig3} are:
i) there is a longer tail of the correlation distributions extending toward lower values for both tips,
resulting in much lower minimum correlations (e.g., 0.26 for the $\mathrm{W_{sharp}}$ tip at positive bias voltages and
0.58 for the $\mathrm{W_{blunt}}$ tip at all bias ranges), ii) the maximum correlations are increased to 0.99 at negative bias for
both tips, iii) the difference between the two distinct peaks of the correlation distributions for the negative and positive bias
in case of the $\mathrm{W_{sharp}}$ tip is reduced, but still significant (above 0.1).

\begin{figure*}
\includegraphics[width=1.00\textwidth,angle=0]{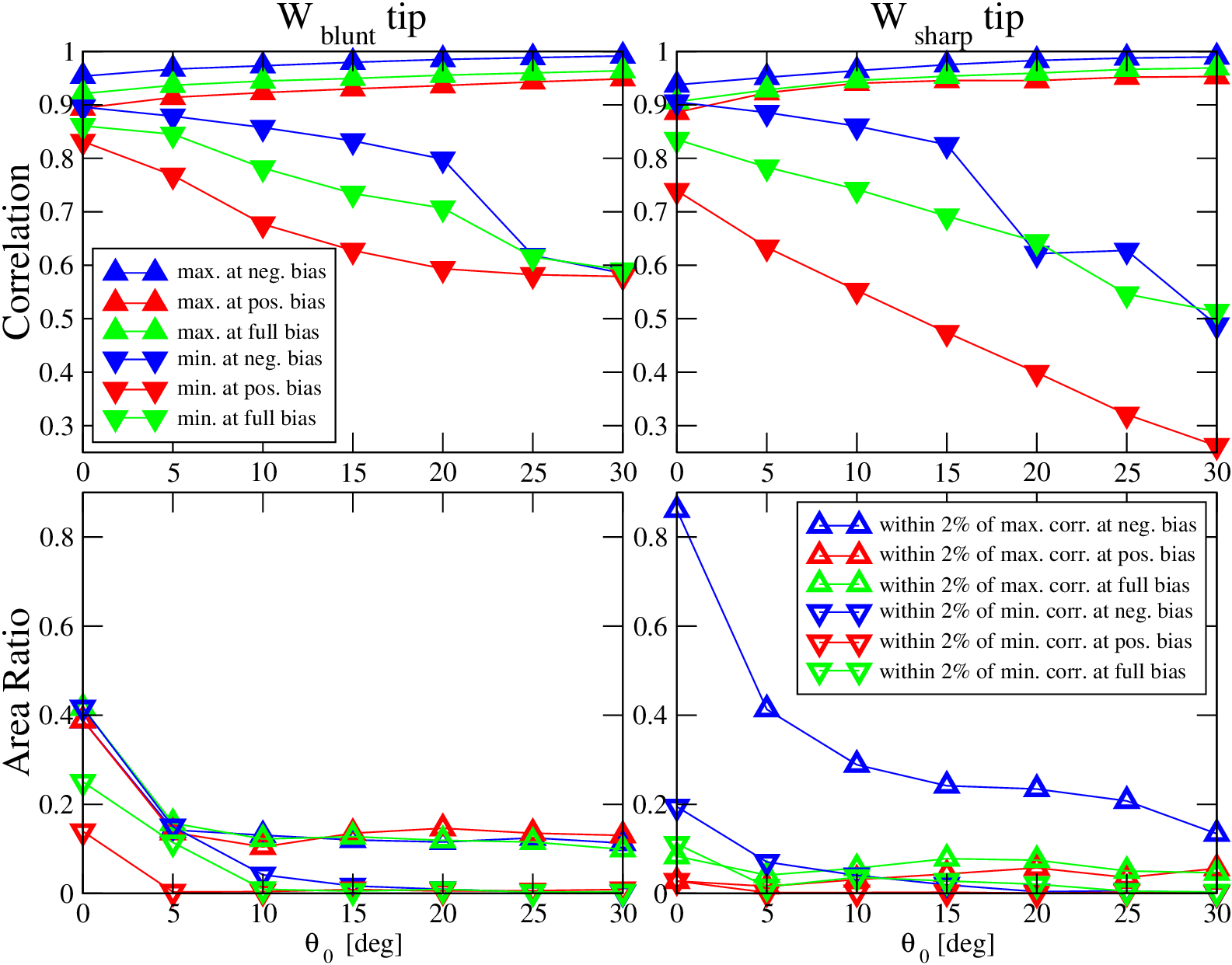}
\caption{\label{Fig9} Extracted data from the correlation maps in the $|V|\le 0.3$ V bias voltage range.
Top row: The evolution of the maximum and minimum correlation value in the $r(\theta_0,\phi_0,\psi_0)$ maps with $\theta_0$.
Bottom row: The $\theta_0$-evolution of the area ratio, for explanation see the caption of Fig.\ \ref{Fig7}.
Left and right parts respectively correspond to data obtained by $\mathrm{W_{blunt}}$ and $\mathrm{W_{sharp}}$ tip models.
}
\end{figure*}

Fig.\ \ref{Fig9} shows the evolutions of the maximum and minimum correlation values and the calculated area ratios with $\theta_0$
obtained from the $r(\theta_0,\phi_0,\psi_0)$ correlation maps in the $|V|\le 0.3$ V bias voltage range. We find that
the main discussed tendencies in Fig.\ \ref{Fig7} are not affected in the low bias regime. However, the area ratios within 2\% of
the maximum correlation are systematically larger for the $\mathrm{W_{sharp}}$ than for the $\mathrm{W_{blunt}}$ tip in the
negative bias range. Since the maximum correlations are above 0.93 for for both type of tips in this bias interval, this
suggests that more tip orientations of the $\mathrm{W_{sharp}}$ tip result in better agreement with the experiment than of the
$\mathrm{W_{blunt}}$ tip at low negative bias, -0.3 V $\le V<$ 0 V. The indications of a favored $\mathrm{W_{blunt}}$ tip in the
experiment are, however, not affected in the other considered low bias regimes.

Although using larger bias ranges is better for the statistical analysis, the tip may become unstable in the experiment at
larger bias voltages, thus making the assignment of the tip geometry and orientation more difficult.
In general, we suggest that the primary decision for the quality of the STM tip in an experiment has to be based on the comparison
between the maximum and minimum relative brightness correlations between two (or more) tip models, and the secondary decisive
factor should be the introduced area ratio measure that gives information on the number of likely or unlikely tip orientations.

\section{Conclusions}
\label{sec_conclusions}

In scanning probe experiments the scanning tip is the source of one of the largest uncertainty as very little is
known about its precise atomic structure and stability. Since the atomic structure and electronic properties of the tip apex can
strongly affect the contrast of STM images, it is very difficult to experimentally obtain predictive
STM images in certain systems. To tackle this problem
we proposed a statistical correlation analysis method to obtain information on the local geometry and orientation of the tip used
in STM experiments. We defined the relative brightness correlation of constant-current topographs between experimental and
simulated data, and analyzed it statistically for the HOPG(0001) surface in combination with two tungsten tip geometries in 18144
orientations. The simulations were performed using the 3D-WKB electron tunneling theory based on first principles electronic
structure calculations. We find that a blunt tip model provides better correlation with the experiment for a wider range of
tip orientations and bias voltages than a sharp tip model. A favored sharp tip is indicated at low negative bias only.
From the correlation distribution we proposed particular tip orientations that are most likely present in the STM experiment, and
likely excluded other orientations. Importantly, we find that the favored relative tip-sample orientations do not correspond to
high symmetry setups that are routinely used in standard STM simulations. The demonstrated combination of large scale simulations 
with experiments is expected to open up the way for a more reliable interpretation of STM data in the view of local tip geometry
effects. Moreover, the introduced correlation analysis method could be useful for other scanning probe imaging techniques as well.

\section*{Acknowledgments}

The authors thank E. Inami, J. Kanasaki, and K. Tanimura at Osaka University for the experimental brightness data.
Financial support of the Magyary Foundation, EEA and Norway Grants, the Hungarian Scientific Research Fund project OTKA PD83353,
the Bolyai Research Grant of the Hungarian Academy of Sciences, and the New Sz\'echenyi Plan of Hungary
(Project ID: T\'AMOP-4.2.2.B-10/1--2010-0009) is gratefully acknowledged. G. T. is supported by EPSRC-UK (EP/I004483/1).
Usage of the computing facilities of the Wigner Research Centre for Physics and the BME HPC Cluster is kindly acknowledged.

\newpage

\appendix
\section*{Appendix: 3D-WKB tunneling theory}
\setcounter{section}{1}
\label{sec_theory}

M\'andi {\it{et al.}} have developed an orbital-dependent electron tunneling model with arbitrary tip orientations
\cite{mandi13tiprot} for simulating scanning tunneling microscopy (STM) measurements within the three-dimensional (3D)
Wentzel-Kramers-Brillouin (WKB) framework based on previous atom-superposition theories
\cite{palotas12orb,palotas11stm,palotas12sts,tersoff85,yang02,smith04,heinze06}.
Here, this method is briefly described, which was used in the paper for the HOPG(0001) surface in combination with tungsten tips.
The model assumes that electrons tunnel through a tip apex structure consisting of a few atoms, and transitions between individual
atoms of this tip apex structure and a suitable number of sample surface atoms, each described by the one-dimensional (1D) WKB
approximation, are superimposed \cite{palotas12orb,palotas11sts}. Since the 3D geometry of the tunnel junction is considered,
the method is a 3D-WKB atom-superposition approach. The advantages, particularly computational efficiency, limitations, and the
potential of the 3D-WKB method were discussed in Ref.\ \cite{palotas13fop}.

The electronic structure of the surface and the tip is included in the model by taking the atom-projected electron density of
states (PDOS) obtained by {\it{ab initio}} electronic structure calculations \cite{palotas11stm}. The orbital-decomposition of the
PDOS is necessary for the description of the orbital-dependent electron tunneling \cite{palotas12orb}.
The energy-dependent orbital-decomposed PDOS functions of the $i$th sample surface atom with orbital symmetry $\sigma$ and the
$j$th tip atom with orbital symmetry $\tau$ are denoted by $n_{S\sigma}^i(E)$ and $n_{T\tau}^j(E)$, respectively.
In the present work $\sigma\in\{s,p_y,p_z,p_x\}$ atomic orbitals for the carbon atoms on the HOPG surface and
$\tau\in\{s,p_y,p_z,p_x,d_{xy},d_{yz},d_{3z^2-r^2},d_{xz},d_{x^2-y^2}\}$ orbitals for the apex atoms of blunt and sharp tungsten
tips are considered. The total PDOS function is the sum of the orbital-decomposed contributions:
\begin{equation}
n_S^i(E)=\sum_{\sigma}n_{S\sigma}^i(E),
\end{equation}
\begin{equation}
n_T^j(E)=\sum_{\tau}n_{T\tau}^j(E).
\end{equation}
Note that a similar decomposition of the Green's functions was reported within the linear combination of atomic orbitals (LCAO)
framework in Ref.\ \cite{mingo96}.

Assuming elastic electron tunneling at temperature $T=0$ K, the tunneling current at the tip apex position $\mathbf{R}_{TIP}$
and bias voltage $V$ is given by the superposition of atomic contributions from the sample surface (sum over $i$),
superposition of atomic contributions from the tip apex structure (sum over $j$) and the
superposition of transitions from all atomic orbital combinations between the sample and the tip (sum over $\sigma$ and $\tau$):
\begin{equation}
\label{Eq_current}
I\left(\mathbf{R}_{TIP},V\right)=\sum_i\sum_j\sum_{\sigma,\tau}I_{\sigma\tau}^{ij}\left(\mathbf{R}_{TIP},V\right).
\end{equation}
One particular current contribution can be calculated as an integral in an energy window corresponding to the bias voltage $V$ as
\begin{eqnarray}
I_{\sigma\tau}^{ij}\left(\mathbf{R}_{TIP},V\right)&=&\epsilon^2\frac{e^2}{h}\int_0^V T_{\sigma\tau}\left(E_F^S+eU,V,\mathbf{d}_{ij}\right)\nonumber\\
&\times&n_{S\sigma}^i\left(E_F^S+eU\right)n_{T\tau}^j\left(E_F^T+eU-eV\right)dU.
\label{Eq_current_decomp}
\end{eqnarray}
Here, $e$ is the elementary charge, $h$ is the Planck constant, and $E_F^S$ and $E_F^T$ are the Fermi energies of the sample
surface and the tip, respectively. The $\epsilon^{2}e^{2}/h$ factor ensures the correct dimension of the electric current.
The value of $\epsilon$ has to be determined by comparing the simulation results with experiments, or with
calculations using standard methods, e.g., the Bardeen approach \cite{bardeen61}. In our simulations $\epsilon=1$ eV was chosen
that gives comparable current values with those obtained by the Bardeen method \cite{palotas12orb} implemented in the BSKAN code
\cite{hofer03pssci,palotas05}. Note that the choice of $\epsilon$ has no qualitative influence on the reported results, and
a rigorous comparison between the 3D-WKB and Bardeen methods in relation to STM experiments performed on HOPG \cite{teobaldi12}
was reported in Ref.\ \cite{mandi14rothopg}.

In Eq.(\ref{Eq_current_decomp}), $T_{\sigma\tau}\left(E,V,\mathbf{d}_{ij}\right)$ is the orbital-dependent tunneling transmission
function, and it gives the probability of the electron tunneling from the $\tau$ orbital of the $j$th tip atom to the
$\sigma$ orbital of the $i$th surface atom, or vice versa, depending on the sign of the bias voltage. The conventions of tip
$\rightarrow$ sample tunneling at positive bias voltage ($V>0$) and sample $\rightarrow$ tip tunneling at negative bias ($V<0$)
are used.
The transmission probability depends on the energy of the electron ($E$), the bias voltage ($V$), and the relative position of the
$j$th tip atom and the $i$th sample surface atom ($\mathbf{d}_{ij}=\mathbf{R}_{TIP}^j-\mathbf{R}_i$). Note that $\mathbf{R}_{TIP}$
corresponds to the position of the tip apex atom. The following form for the transmission function is considered
\cite{mandi13tiprot}:
\begin{equation}
\label{Eq_Transmission}
T_{\sigma\tau}\left(E_F^S+eU,V,\mathbf{d}_{ij}\right)=\exp\{-2\kappa(U,V)|\mathbf{d}_{ij}|\}\chi_{\sigma}^2(\theta_{ij},\phi_{ij})\chi_{\tau}^2(\theta_{ij}',\phi_{ij}').
\end{equation}
Here, the exponential factor corresponds to an orbital-independent transmission, where all electron states are considered as
exponentially decaying spherical states \cite{tersoff85,heinze06,tersoff83}, and it depends on the distance between the $j$th tip
atom and the $i$th surface atom, $|\mathbf{d}_{ij}|$, and on the vacuum decay,
\begin{equation}
\label{Eq_kappa}
\kappa(U,V)=\frac{1}{\hbar}\sqrt{2m\left(\frac{\varphi_S+\varphi_T+eV}{2}-eU\right)}.
\end{equation}
For using this $\kappa$ an effective rectangular potential barrier in the vacuum between the sample and the tip is assumed.
$\varphi_S$ and $\varphi_T$ are the electron work functions of the sample surface and the tip, respectively, $m$ is the electron's
mass and $\hbar$ the reduced Planck constant. The remaining factors of Eq.(\ref{Eq_Transmission}) are responsible for the
orbital dependence of the transmission. They modify the exponentially decaying part according to the real-space shape of the
electron orbitals involved in the tunneling, i.e., the angular dependence of the electron densities of the atomic orbitals of
the surface and the tip is taken into account as the square of the real spherical harmonics
$\chi_{\sigma}(\theta_{ij},\phi_{ij})$ and $\chi_{\tau}(\theta_{ij}',\phi_{ij}')$, respectively. It is important to note that
the angles are given in the respective local coordinate system of the surface (without primes) and the tip
(denoted by primes). This distinction of the local coordinate systems is crucial to describe arbitrary tip orientations that
correspond to a rotation of the tip coordinate system by the set of Euler angles $(\theta_0,\phi_0,\psi_0)$ with respect
to the surface coordinate system \cite{mandi13tiprot}. The transformation between a vector defined in the local coordinate system
of the tip, $(x',y',z')$, and a vector defined in the local coordinate system of the sample, $(x,y,z)$, is given by
\begin{equation}
\left(
\begin{array}{c}
x'\\
y'\\
z'
\end{array}
\right)
=\underline{\underline{R}}(\theta_0,\phi_0,\psi_0)
\left(
\begin{array}{c}
x\\
y\\
z
\end{array}
\right),
\end{equation}
with the rotation matrix:
{\footnotesize{
\begin{eqnarray}
\label{Eq_mrot}
&&\underline{\underline{R}}(\theta_0,\phi_0,\psi_0)=\\
&&
\left(
\begin{array}{ccc}
\cos\phi_0\cos\psi_0-\sin\phi_0\sin\psi_0\cos\theta_0 & \cos\phi_0\sin\psi_0+\sin\phi_0\cos\psi_0\cos\theta_0 & \sin\phi_0\sin\theta_0\\
-\sin\phi_0\cos\psi_0-\cos\phi_0\sin\psi_0\cos\theta_0 & -\sin\phi_0\sin\psi_0+\cos\phi_0\cos\psi_0\cos\theta_0 & \cos\phi_0\sin\theta_0\\
\sin\psi_0\sin\theta_0 & -\cos\psi_0\sin\theta_0 & \cos\theta_0
\end{array}
\right).\nonumber
\end{eqnarray}
}}
The polar and azimuthal angles $(\theta_{ij}^{(')},\phi_{ij}^{(')})$ given in both real spherical harmonics in
Eq.(\ref{Eq_Transmission}) correspond to the tunneling direction, i.e., the line connecting the $i$th surface atom and the
$j$th tip atom, as viewed from their local coordinate systems (denoted by no prime and prime, respectively), and they have to be
determined for each surface atom-tip atom ($i-j$) combination from the actual tip-sample geometry. A schematic view of an STM
tip with rotated local coordinate system by the Euler angles $(\theta_0,\phi_0,\psi_0)$ above the HOPG(0001) surface is shown
in Fig.\ \ref{Fig1}. $\theta$, $\phi$ and $d$ are also shown for a given $i-j$ pair. For more details of the 3D-WKB formalism,
see Refs.\ \cite{mandi13tiprot,palotas12orb}, and for a rigorous comparison between the 3D-WKB and Bardeen methods in relation
to STM experiments performed on HOPG, see Ref.\ \cite{mandi14rothopg}.

\end{document}